%% file: emergent-gravity.tex
\documentclass[a4paper,11pt]{article}
\pdfoutput=1 

\usepackage{jheppub} 

\usepackage{url}
\usepackage{amsmath}
\usepackage{amsthm}
\usepackage{amssymb}
\usepackage{microtype}
\usepackage{bm}
\usepackage{graphicx}
\usepackage{color}

\theoremstyle{plain}
\newtheorem{thm}{Theorem}[section]

\theoremstyle{definition} 
\newtheorem{defn}[thm]{Definition}

\theoremstyle{remark}
\newtheorem{remark}[thm]{Remark}

\numberwithin{equation}{section}

\input{emer-grav-defs}

\title{\boldmath Emergent Gravity and Weyl's Volume Formula}
\author[1]{Orlando Alvarez%
\note{This work was supported in part by the
National Science Foundation under Grant PHY-1212337.}}
\affiliation{Department of Physics, University of Miami,\\
1320 Campo Sano Ave, Coral Gables, FL 33146 USA}
\emailAdd{oalvarez@miami.edu}

\abstract{In physical theories where the energy (action) is localized
near a submanifold of Euclidean (Minkowski) space, there is a
universal expression for the energy (or the action).  We derive a
multipole expansion for the energy that has a finite number of terms,
and depends on intrinsic geometric invariants of the submanifold and
extrinsic invariants of the embedding of the submanifold.  This
universal expression is a generalization of an exact formula of
Hermann Weyl for the volume of a tube.  We describe when our result is
applicable, when our generalization gives an exact result, and when
there are corrections (often exponentially small) to our formula.  In
special situations, dictated by spherical symmetry, the expression is
a generalized Lovelock lagrangian for gravity, a class of theories
that are interesting because they have no negative metric states.  We
discuss whether these results represent a true theory of emergent
gravity by discussing simple models where a higher dimensional quantum
field theory without a fundamental graviton leads to a gravity-like
theory on a submanifold where all or some of the dynamical degrees of
freedom are fluctuations of the metric on the submanifold.
}%

\begin{document}
\maketitle
\flushbottom

\section{Introduction}
\label{sec:intro}

We consider physical systems where there is a localization of energy
(or action) near a submanifold $\Sigma$ of Euclidean space $\bbE^{n}$
(or Minkowski space $\bbM^{n}$).  What we argue in this article is
that there is a universal expression for the energy (or
the action), and we derive a multipole expansion for the energy that
has a finite number of terms, and depends on intrinsic geometric
invariants of the submanifold and extrinsic invariants of the
embedding of the submanifold.  We briefly discuss two potential
examples.

\begin{figure}[tbp]
    \centering
    \includegraphics[width=0.7\textwidth]{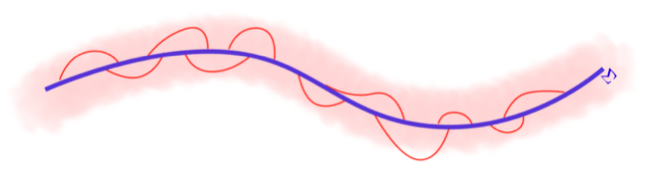}
    \caption[xyz3967]{D-brane $\Sigma$ with open strings terminating
    on it.  The shaded volume denotes an effective energy density due
    to the presence of the open strings.  The string tension
    $(2\pi\alpha')^{-1}$ keeps the strings from getting too long and
    this effectively sets the radius of the energy localization near 
    the brane.}
    \protect\label{fig:brane}
\end{figure}
First, consider a $p$-dimensional D-brane \cite{Dai:1989ua} whose time
evolution is a $q= p+1$ dimensional submanifold\footnote{We write
$\Sigma^{q}$ when we want to be explicit about $\dim\Sigma$.} $\Sigma=
\Sigma^{q}$ of $n$-dimensional Minkowski space $\bbM^{n}$.  Open
strings of length $O(\sqrt{\alpha'})$, where $\alpha'$ is the Regge
slope, terminate on a $\Sigma$ as in Figure~\ref{fig:brane}.  The open
strings arise due to quantum fluctuations and this mechanism leads to
an effective action density localized within a distance of
$O(\sqrt{\alpha'})$ near the submanifold $\Sigma$.  In a string
theory, there are closed strings propagating in the bulk, \emph{i.e.},
the string theory has a bulk graviton.  Theories of gravity of this
type in a Kaluza-Klein compactification scenario were discussed in
\cite{Antoniadis:1998ig}.

A second example is the effective action for a topological defect 
\cite{Forster:1974ga,Maeda:1987pd,Gregory:1988qv}.  For
concreteness consider the Nielsen-Olesen vortex in the $\U(1)$
Higgs model in $\bbM^{4}$.  The surface $\Sigma$
corresponds to the $2$-dimensional timelike submanifold that is the locus of
$\phi=0$ where $\phi$ is the complex scalar field. The scalar field 
and the gauge field decay exponentially as you move away from 
$\Sigma$ therefore the action density of the theory also decays 
exponentially as you move orthogonally to $\Sigma$.

The use of geometric concepts in computing the energy of a physical
system has a long history in physics.  For example, the very
successful phenomenological liquid drop model in nuclear physics uses
the short range nature of the nuclear force and the incompressibility
of nucleons to argue that the binding energy of a nucleus with atomic
number $A$ goes like $a A - b A^{2/3} + \dotsb$.  The first term is a
volume energy term and the second term is a surface area term
reflecting that the nucleons on the boundary have less neighbors for
interaction.

In this article we derive a general expansion for models with
localized energy near a submanifold in terms of the curvature
invariants of the submanifold in question.  The motivation comes from
an formula due to Hermann Weyl for the volume of a
tube~\cite{Weyl:tubes}.
In the simplest models we will show that the dynamics of the 
theory is determined by an effective action with a finite number of 
terms with universal form:
\begin{equation}
    \begin{split}
	I & = \int_{\Sigma}\Phi_{0}(\sigma)\; \sqrt{-\det g}\; d^{q}\sigma
	+ \int_{\Sigma}\Phi_{1}(\sigma) \;R\sqrt{-\det g}\;
	d^{q}\sigma \\
	&\quad + \int_{\Sigma}\Phi_{2}(\sigma) \left( R^{2} - 4
	R_{ab}R^{ab} + R_{abcd}R^{abcd} \right) \sqrt{-\det g}\;
	d^{q}\sigma \\
	&\qquad + \text{[a finite number of known terms, see
	eq.~\eqref{eq:gen-Weyl-x}]},
    \end{split}
    \label{eq:eff-1}
\end{equation}
where $g_{ab}$ is the metric on $\Sigma^{q}$, $R_{abcd}$ is the
Riemannian curvature tensor of that metric, and $\Phi_{0}(\sigma)$,
$\Phi_{1}(\sigma)$ and $\Phi_{2}(\sigma)$ are effective fields on
$\Sigma$.  In the simplest models for topological defects, the values
of the $\Phi$ fields are constants.  In D-brane models, the field
$\Phi_{0}$ is the Dirac-Born-Infeld (DBI) Lagrangian, see 
Section~\ref{sec:DBI}.

The terms that appear to all orders in \eqref{eq:eff-1} are the
``dimensional continuations'' of the Euler densities known as the
Lipschitz-Killing curvatures\footnote{Nineteenth Century
mathematicians, R.~Lipschitz and W.~Killing were interested in
invariant theory and in geometry.  They wrote down these curvature
expressions as geometrical invariants of a manifold without any
inkling of the relationship to  the topological invariants discovered
by S.S.~Chern much later.}.  From the physics viewpoint this is
astonishing.  Gravitational theories defined by Lagrangians containing
those terms were discussed by Lovelock
\cite{Lovelock:1971yv,Lovelock:1972vz} in the early 1970s who was
interested constructing generalizations of the Einstein tensor.  He
required his tensors to be symmetric, rank two, divergence-free and
to contain at most the first two derivatives of the metric.
The appearance of Lovelock Lagrangians in string theory was first
observed by Zwiebach \cite{Zwiebach:1985uq} who noted that
compatibility of a ghost free theory with the presence of curvature
squared terms in the gravitational Lagrangian required a special
combination that reduced to the Euler density in four dimensions.  By
studying the $3$-graviton on shell vertex in string theory he verified
that this curvature squared combination appears.  Zumino
\cite{Zumino:1985dp} generalized Zwiebach's results and showed that
gravitational theories containing higher powers of the curvature given
by a Lagrangian where the additional terms were ``dimensional
continuations'' of Euler densities in the appropriate dimensionality,
\emph{i.e.}, Lovelock type Lagrangians, were ghost free.  A Lovelock
gravitational action for a spacetime $\Sigma^{q}$ is constructed by
taking linear combinations of specific curvature invariants:
\begin{equation}
    I = \sum_{r=0}^{\lfloor q/2 \rfloor} \lambda_{2r}\; I_{2r} =
    \sum_{r=0}^{\lfloor q/2 \rfloor} \lambda_{2r}\; \int_{\Sigma}
    \curv_{2r}\; \dual_{\Sigma}\,,
    \label{eq:Lovelock-lag-0}
\end{equation}
where $\lambda_{2r}$ is a coupling constant, $\dual_{\Sigma}$ is the
volume element on $\Sigma$, and the curvature terms $\curv_{2r}$ are
defined by \eqref{eq:def-curv-x}.  The contribution to the equations
of motion from each action $I_{2r}$ is a symmetric divergenceless
second rank tensor that contains at most second derivatives of the
metric \cite{Lovelock:1971yv,Zumino:1985dp}.  These Lovelock
Lagrangian densities arise in the context of this article because of a
formula~\cite{Weyl:tubes} due to Hermann Weyl for the volume of a
\emph{tube} that we describe beginning in
Section~\ref{sec:physics-tubes}. What we derive here is a 
generalization of Weyl's formula.

We mostly work in Euclidean signature and use the language of
statistical mechanics or of Euclidean quantum field theory.  We are interested
in the isometric embedding of a $q$-dimensional submanifold $\Sigma =
\Sigma^{q}$ of Euclidean space $\bbE^{n}$.  If the embedding\footnote{We use a hooked
arrow $\hookrightarrow$ to denote an embedding.} is described by a map
$X: \Stil \to \bbE^{n}$ with $\Sigma = X(\Stil) \hookrightarrow
\bbE^{n}$ then the metric induced from the embedding is
\begin{equation}
    g_{ab}(\sigma)\; d\sigma^{a} \otimes d\sigma^{b} = \delta_{\mu\nu}\; \frac{\partial 
    X^{\mu}}{\partial \sigma^{a}}\; \frac{\partial X^{\nu}}{\partial 
    \sigma^{b}} \; d\sigma^{a} \otimes d\sigma^{b}\;,
    \label{eq:def-g}
\end{equation}
and the induced Euclidean volume is 
\begin{equation}
    \vol_{q}(\Sigma) = \int_{\Sigma}
    \sqrt{\det g}\; d^{q}\sigma\;.
    \label{eq:def-vol}
\end{equation}
The volume is invariant under the action of the Euclidean group on the
embedding map $X:\Stil \hookrightarrow \bbE^{n}$ and under the action of
$\Diff_{0}(\Stil)$, the group of diffeomorphisms of $\Stil$
connected to the identity.  Note that the induced metric $g$ can be
determined intrinsically via measurements performed on the surface.

We point out some related research that is not the main focus of this
work.  In our work we start with a higher dimensional theory without
gravity and we obtain a gravity-like theory on a submanifold.  This
is dual to the work of
Maldacena~\cite{Maldacena:1997re,Aharony:1999ti} where in the bulk he
begins with a theory of gravity and on the conformal boundary he has a
Yang-Mills theory. We are not aware of correlation function relations 
as in the Maldacena duality.

In some sense, the most elegant theories of emergent gravity are 
topological such as the A and B topological theories
\cite{Witten:1988xj,Witten:1988ze,Bershadsky:1993cx}.  We do not see a
direct connection of our work to these since we begin in a metric
space $\bbE^{n}$ or $\bbM^{n}$.

Our work is also related to theories for emergent gravity or gravity
on submanifolds.  There are many scenarios and our research touches
three major areas.  There is a lot of work on models to explain the
relative weakness of the gravitational force in four dimensions
starting with a theory of gravity in higher dimensions.  They are the
Kaluza-Klein type models of Antoniadis, Arkani-Hamed, Dimopoulos and
Dvali~\cite{ArkaniHamed:1998nn,ArkaniHamed:1998rs,Antoniadis:1998ig}
(AADV) and their variants where a TeV scale higher dimensional gravity
with a compactification scale large compared to the Planck length
leads to a four dimensional gravity theory with the experimental value
for the Planck mass $\MPl_{4}\sim 10^{19}~\text{TeV}$.  There are also
the models motivated by the work of Randall and
Sundrum~\cite{Randall:1999vf,Randall:1999ee}.  Here the universe is a
slice of $\AdS_{5}$ with two boundary pieces of which one is the
physical $3$-brane we inhabit.  The radius of curvature of the
$\AdS_{5}$ and width of the slice are Planck length order of
magnitude.  Randall and Sundrum use the exponential change in the
metric as you move from the hidden slice to the visible slice to
generate an exponential hierarchy where the physical field theoretic
parameters on the visible slice are TeV scale.  An extension of the
work presented here to the case of embedding in constant curvature
spaces and the Randall-Sundrum scenario appears in a companion
article~\cite{Alv:2016b}.  There are also the models of emergent
gravity motivated by the work of Dvali, Gabadadze and
Porrati~\cite{Dvali:2000hr} where gravity on an infinite five
dimensional spacetime reproduces the correct crossover to 4
dimensional behavior on a 4 dimensional submanifold.  The focus of
this paper is theories that do not contain gravity in the bulk but
lead to a gravity-like theory on a submanifold.  A phenomenon like
this happens in theories of defects but we explain carefully in
Sections~\ref{sec:embedding} and \ref{sec:defects} that these do not
necessarily lead to a theory of emergent gravity.

\section{The classical physics of tubes}
\label{sec:physics-tubes}

We first define a \emph{tube} and subsequently discuss the physics of 
tubes. An Euclidean tube, in the sense of Hermann Weyl's classic
paper~\cite{Weyl:tubes}, is a way of thickening a submanifold $\Sigma$
of Euclidean space $\bbE^{n}$. 
For example, the thickening of a point will be a small ball.  An
obvious thickening of a line in $\bbE^{3}$ is a solid cylinder of
small radius.  A natural thickening of the $q$-dimensional submanifold
$\Sigma^{q}$ in $\bbE^{n}$ is the $n$-dimensional \emph{tube},
$\tube(\Sigma^{q},\rho)$, of radius $\rho$ constructed from $\Sigma$ by
thickening the submanifold by moving orthogonally to the submanifold a
distance $\rho$, see Figure~\ref{fig:tube}.  The technical
mathematical definition~\cite{Gray:tubes} is given below.
\begin{figure}[tbp]
    \centering
    \includegraphics[width=0.4\textwidth]{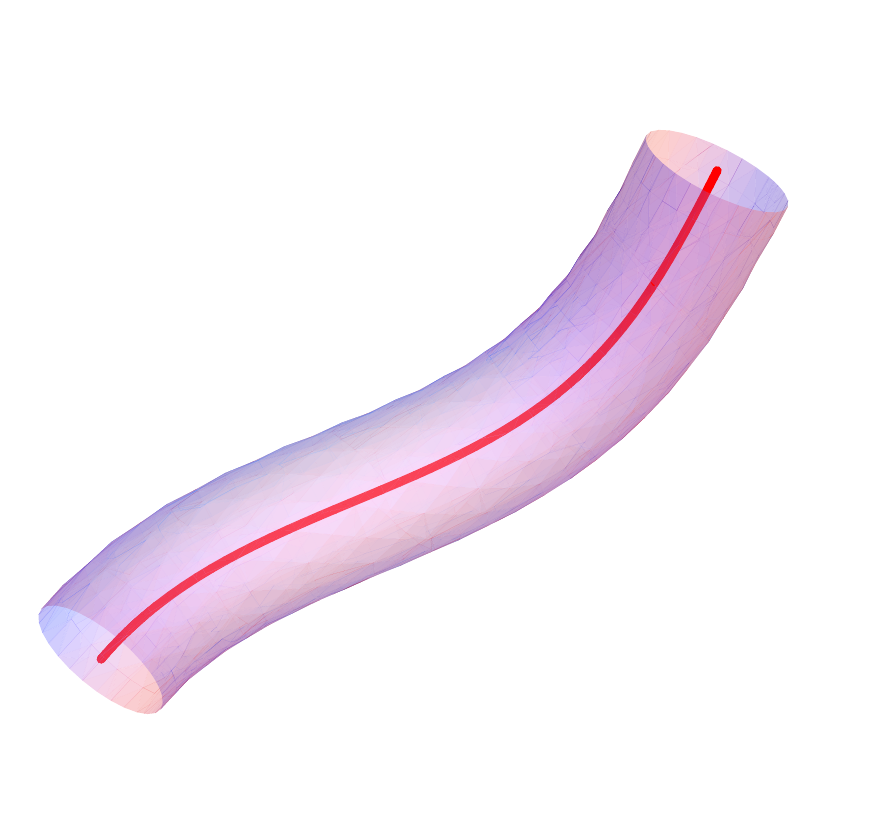}
    \caption[xxx1]{This is a portion of a tube corresponding to
    the thickening of a closed curve (the red curve) in
    $\bbE^{3}$.  The boundary of the tube is a fixed distance
    $\rho$ from the curve.  
    }
    \label{fig:tube}
\end{figure}

\begin{defn}[\textsf{Weyl Tube}]
     Let $\Sigma^{q} \hookrightarrow \bbE^{n}$ be an embedded compact
     submanifold without boundary, \emph{i.e.} a closed submanifold.
     The \emph{tube} $\tube(\Sigma^{q},\rho)$ of radius $\rho$ about
     $\Sigma^{q}$ is a subset of $\bbE^{n}$ with the following
     characterization: $\bx \in \bbE^{n}$ is in the tube if there
     exists a straight segment from $\bx$ to $\Sigma^{q}$ that intersects
     $\Sigma^{q}$ perpendicularly and the length of the segment is less
     than or equal to $\rho$, see Figure~\ref{fig:tube}.  The tube
     $\tube(\Sigma^{q},\rho)$ is a fiber bundle over $\Sigma^{q}$ with fiber
     $B^{l}$, the $l$-dimensional ball (the solid $(l-1)$-sphere).
\end{defn}

\begin{remark}
    In this article we  relax the requirements on the definition 
    of a tube by allowing submanifolds that are not compact but have no 
    boundary such as a $q$-dimensional vector subspace of $\bbE^{n}$.
\end{remark}

\begin{figure}[tbp]
    \centering
    \includegraphics[width=0.55\textwidth]{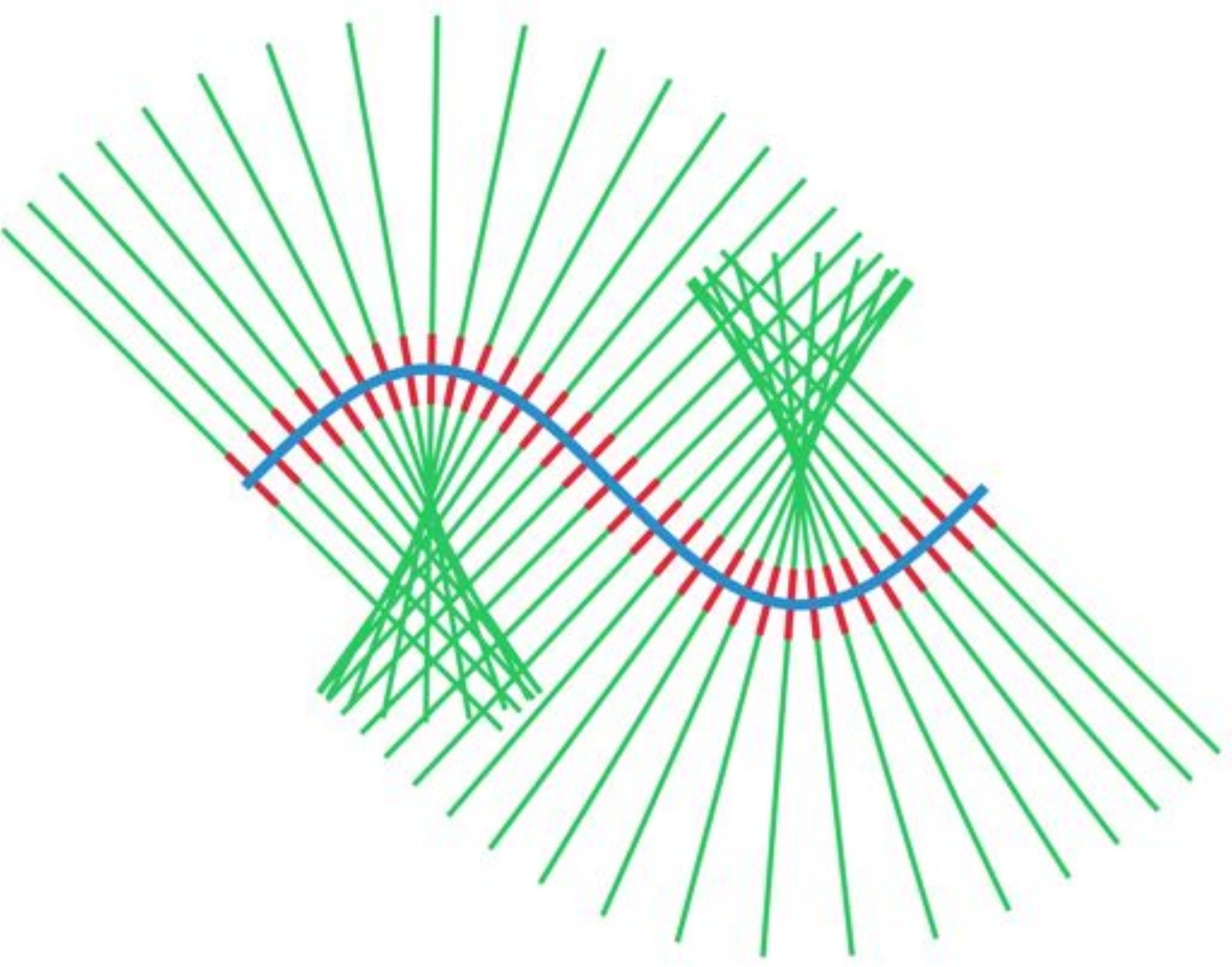}
    \caption[xyz353]{An example of what happens if the radius of the
    tube is made too big.  A small radius tube (red normal
    neighborhood) is okay but a large radius tube (green normal
    neighborhood) could have self intersections.  Locally, if the
    radius $\rho$ of the tube is small compared to the radius of
    curvature of $\Sigma$, roughly in the sense that $\rho \lVert
    \bK_{\Sigma} \rVert < 1$, then there will not be any self
    intersections.  Here $\bK_{\Sigma}$ is the extrinsic curvature
    tensor of $\Sigma$, \emph{i.e.}, the second fundamental form.
    Imposing the much stronger requirement, $\rho \lVert \bK_{\Sigma}
    \rVert \ll 1$, is too draconian and leaves out potentially
    interesting physics applications.  This is a public domain image
    from
    \href{http://commons.wikimedia.org/wiki/File:Tubular_neighborhood.png}%
    {Wikipedia Commons}.}
    \label{fig:tube-large-radius}
\end{figure}

\begin{figure}[tbp]
    \centering
    \includegraphics[width=0.35\textwidth]{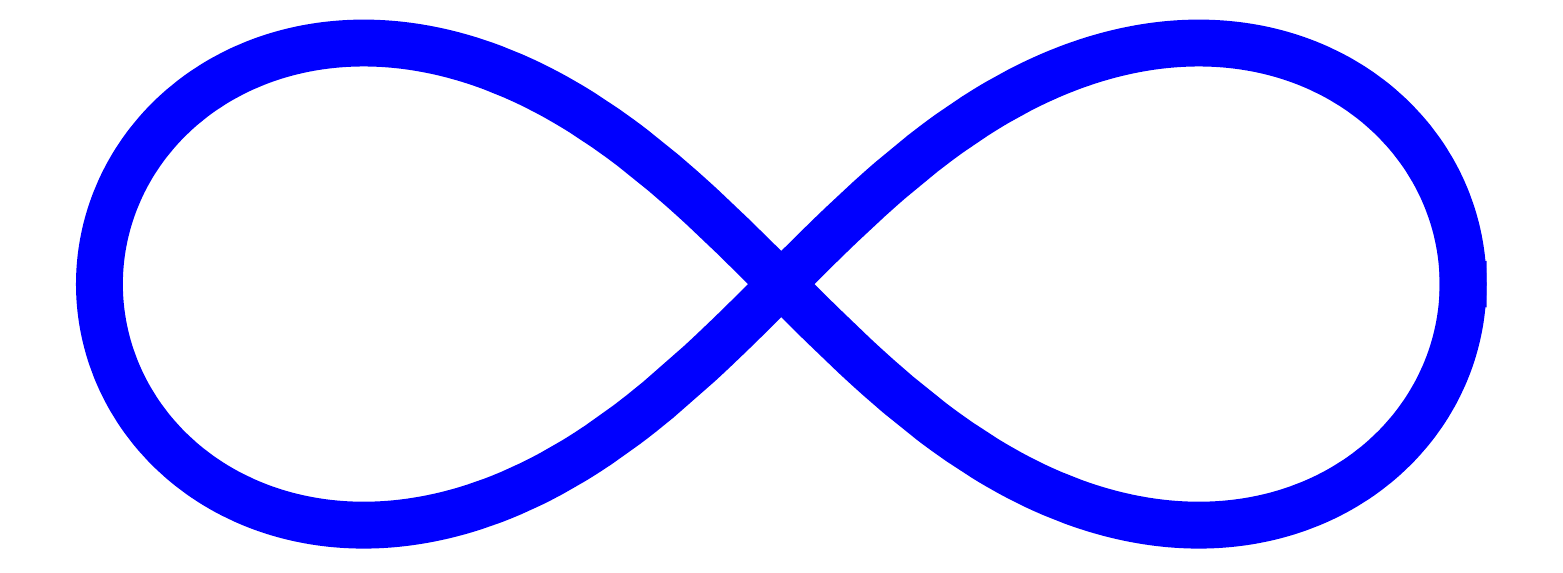}
    \caption[xyz583]{Here we have thickened a planar lemniscate of
    Bernoulli that is embedded in $\bbE^{2}$.  Everywhere the radii of
    curvature are larger than the radius of the tube so locally we do
    not expect self intersections.  There are self intersections
    globally because the curve can come close to itself (here the
    lemniscate intersects at the origin) and consequently the tubes
    can intersect.  We will ignore similar situations in this paper.}
    \label{fig:lemniscate}
    
\end{figure}

\begin{figure}[tbp]
    \centering
    \includegraphics[width=0.7\textwidth]{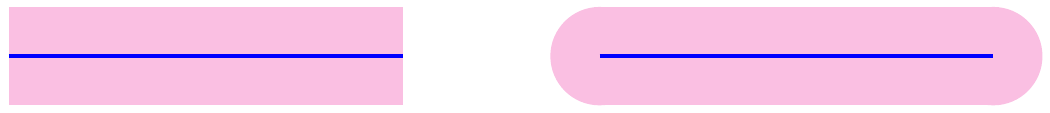}
   \caption[xxx265]{If the submanifold $\Sigma$ has a boundary then
   you have to be careful about what the tube $\tube(\Sigma,\rho)$
   looks like near $\partial\Sigma$.  The figure on the left is a tube
   but the one on the right is not.  Note that all points of the tube
   in both cases are closer than a distance $\rho$ from $\Sigma$ but
   not all shortest distance geodesics are orthogonal to $\Sigma$ in
   the figure on the right.  Weyl's volume formula is not valid for
   the example on the right.}
   \label{fig:tube-boundary}
\end{figure}

Clearly the tube changes as we vary the extrinsic geometry of $\Sigma^{q}$
while keeping the intrinsic geometry fixed\footnote{A cylinder and a
$2$-plane in $\bbE^{3}$ have the same local flat intrinsic geometry
but the extrinsic geometries are very different}.  Throughout this
article we always implicitly assume that the radius $\rho$ of the tube
has to be small enough so that there are no (local) self
intersections.  For example, the extrinsic radii of curvature of
$\Sigma^{q}$ should be large compared to $\rho$, see
Figure~\ref{fig:tube-large-radius}.
Note, the local restrictions on the radius of the tube and the radii
of curvature are not sufficient to guarantee that globally we will not
have self intersections, see Figure~\ref{fig:lemniscate}.  If $\Sigma^{q}$
has a boundary then there are technicalities and we do not study this
case here, see Figure~\ref{fig:tube-boundary} and
Reference~\cite{Gray:tubes}. In this article we only consider 
``nice'' tubes and we avoid all subtleties and technicalities.

We first discuss an artificial and mostly academic physical model: the
classical dynamics of a tube.  When we embed a $n$-submanifold, the
standard action is proportional to the $n$-volume of the submanifold.
A tube $\tube(\Sigma^{q},\rho)$ is an $n$-dimensional submanifold, hence
the classical action should be:
\begin{equation}
    I_{\text{tube}} = T_{n}
    \vol_{n}\left(\tube(\Sigma^{q},\rho)\right)
    \label{eq:action-tube}
\end{equation}
where the tension $T_{n}$ has dimension $[T_{n}] = L^{-n}= M^{n}$.
There are no bulk curvature terms to be included in the action because
the bulk metric is the flat Euclidean metric.  In principle there
could be geometric terms in the action associated with the $(n-1)$
dimensional boundary of the tube but we ignore these.

The dynamics of an $n$-submanifold governed by the volume action is
very general.  Imagine an $n$-ball with boundary waves or the $n$-ball
deforming itself to a cylindrical configuration; there are endless
possibilities.  The dynamics of a tube are much more restrictive.
There is an instructive classical mechanical analogy.  The motion
of $10^{25}$ interacting particles in $\bbE^{3}$ is intractable but
the motion of a rigid body made up of $10^{25}$ particles has a six
dimensional configuration space: three translational degrees of
freedom and three rotational degrees of freedom.  Additionally, the
kinetic part of the action of a rigid body only depends on a small
number of parameters: the total mass and the moment of inertia tensor
of the body.  Two body potentials that are translationally and
rotationally invariant will only contribute a constant to the action
of a rigid body.  An external constant gravitational field will give
an effective potential energy that depends on the position of the
center of mass.  These simplifications lead to an action for a rigid
body with tractable equations of motion.

The restriction to the study of the classical motion of a \emph{tube}
and not to a general $n$-submanifold greatly constrains the dynamics
because the tube is uniquely determined by the underlying base
manifold $\Sigma^{q}$ and the radius $\rho$.  What is surprising is
that the action \eqref{eq:action-tube} may be expressed solely in terms
of the intrinsic geometry of $\Sigma$ because of a celebrated formula
\eqref{eq:Weyl-tube-for} for the $n$-volume of a tube due to Hermann Weyl
\cite{Weyl:tubes}.  This formula has exactly $1+\lfloor
q/2\rfloor$ terms and the first three are:
\begin{align}
    \vol_{n}(\tube(\Sigma^{q},\rho)) & = V_{l}(B^{l})\,\rho^{l}\; 
    \vol_{q}(\Sigma)
    + V_{l}(B^{l})\;\frac{\rho^{l+2}}{2(l+2)}\; \int_{\Sigma}
    R\;\dual_{\Sigma} \nonumber\\
     & \quad +V_{l}(B^{l})\; \frac{\rho^{l+4}}{8 (l+2)(l+4)}\; 
     \int_{\Sigma} \left( R^{2} - 4 R_{ab}R^{ab} + 
     R_{abcd}R^{abcd}\right) \dual_{\Sigma} \nonumber\\
     &\quad + O(\rho^{l+6}) \,.
    \label{eq:Weyl-3}
\end{align}
In the above $l=n-q= \codim \Sigma$, and $V_{l}(B^{l})$ is the
$l$-volume of the unit $l$-ball (see eq.~\eqref{eq:vol-ball}),
$R_{abcd}$ is the Riemann curvature tensor of the induced metric on
$\Sigma$, and $\dual_{\Sigma}$ is the intrinsic volume element of the
induced metric on $\Sigma$.  The dynamics of the $n$-dimensional tube
is determined by an action for the motion of the $q$-dimensional base
$\Sigma$ of the tube.  The terms that appear to all orders in $\rho$
in Weyl's tube volume formula are the ``dimensional continuations'' of
the Euler densities.  We refer the reader back to the Introduction
(Section~\ref{sec:intro}) where the significance of such terms is
discussed.

Firstly, we remark on a couple of surprising results that follow from the
Weyl volume formula.  If we thicken a closed curve $\Sigma^{1}
\hookrightarrow \bbE^{3}$ then the volume of the tube will be $\pi
\rho^{2} L$ where $L$ is the length of the curve.  This result is what
you would expect for a right circular cylinder and is independent of the shape of the
embedded curve.  In the case of a closed surface $\Sigma^{2}
\hookrightarrow \bbE^{3}$, we have the exact result
$\vol_{3}(\tube(\Sigma,\rho)) = 2\rho\, \vol_{2}(\Sigma) +
\frac{4\pi}{3}\, \rho^{3}\; \chi(\Sigma)$. Notice that the first 
summand is the naive volume of a slab of thickness $2\rho$ centered about 
$\Sigma$.

Secondly, a remarkable property of Weyl's formula is that the volume
of the tube depends only on the \emph{intrinsic geometry} of $\Sigma$,
\emph{i.e.}, the induced metric and not the extrinsic curvatures.  The
only features of the embedding that appear in the formula are the
dimensionality of the embedding space (via $l= n-q$) and the radius of
the tube~$\rho$.  The extrinsic curvatures (second fundamental forms)
that characterize the embedding of $\Sigma$ do not appear.

If we combine the Weyl volume formula with \eqref{eq:action-tube} then
we see that the classical action of tubes is governed by the same
action as generalized Lovelock theories of gravitation in
$q$-dimensions.  The first term is a $p$-brane tension $T_{q}$ for the
$p$-brane, $T_{q} \sim T_{n} \rho^{l}$, and the second term in the
expansion is the Einstein-Hilbert action with $q$-dimensional
gravitational constant $G_{q}^{-1} \sim T_{n}\rho^{l+2}$.  Note that
the conventionally defined cosmological constant is $\Lambda_{q} =
T_{q} G_{q} \sim 1/\rho^{2}$ is independent of the tube tension
$T_{n}$.  What we have here is a gravity-like action that
\emph{emerges} from embedding dynamical tubes in Euclidean space.
There may be no fundamental graviton in $\bbE^{n}$ but it is possible
that a graviton on $\Sigma$ emerges because of the fluctuations of the
embedding.  This is in the same sense that phonons emerge from lattice
vibrations.  Whether or not we have an emergent graviton is discussed
in Section~\ref{sec:embedding} and is related to the mathematics of
isometrically embedding submanifolds in Euclidean space.  In the case
that we have a \emph{de facto} graviton, the classical physics of the
$n=(q+l)$ dimensional tube $\tube(\Sigma,\rho)$ is the same as
$q$-dimensional Lovelock gravity for the spacetime $\Sigma$.

\section{Energy tubes}
\label{sec:E-tubes}

Tubes in the sense used by Weyl are mathematically interesting but it
is difficult to envision a physical phenomenon that would lead to
their existence.  In this section we consider a generalization of Weyl
tubes that we will call \emph{energy tubes}. In Minkowski space these
should more properly be called \emph{action tubes}.  We can construct
physical systems that lead to energy tubes and we describe several
possibilities in this section.  An energy (action) tube is a
localization of energy (action) near a submanifold.
\begin{figure}[tbp]
    \centering
    \includegraphics[width=0.4\textwidth]{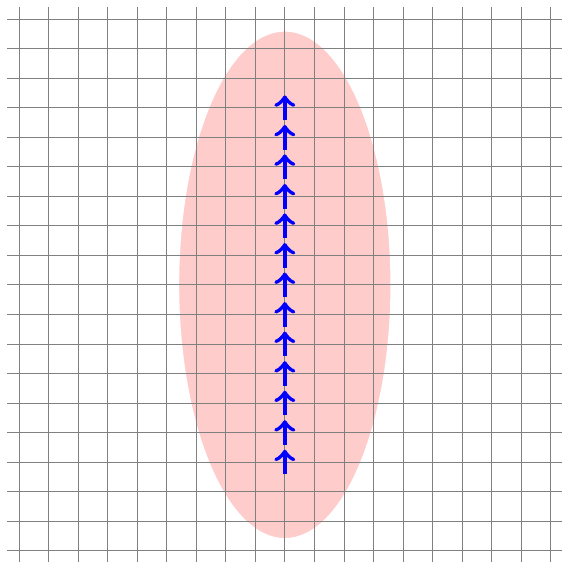}
    \caption[xxx0]{A $2$-d Ising model with a wall of spins fixed to
    point upwards.  Because of the finite range correlations, the
    energy density will differ from the ``bulk energy density'' in a
    small \emph{energy tube} localized near the wall.}
    \label{fig:ising}
\end{figure}    
We motivate the basic idea of an energy tube with a phenomenological
discussion within the $2$-dimensional Ising model.  The Ising
Hamiltonian is $(-1/T) \sum_{\langle \br, \br' \rangle} s(\br)
s(\br')$ where $s(\br)= \pm 1$ is the spin at lattice point $\br$,
$\langle \br, \br' \rangle$ denotes a nearest neighbor pair, and $T$
is the temperature of the heat bath.  Assume we are in the disordered
phase, $T > T_{c}$, with finite correlation length $\xi>0$.  Suitably averaged,
the energy density will be constant throughout the plane.

We now modify the system by introducing a finite ``wall'' where the
spins are all fixed to point up as in Figure~\ref{fig:ising}.  At
distances much larger that the correlation length $\xi$, the
thermodynamic system will be unaware of the wall and the energy
density will be just as in the Ising model without the wall.  Near the
wall, \emph{i.e.}, within several correlation lengths of the wall, the
distribution function for the fluctuating spins will be affected by
the ``wall boundary conditions'' and we expect a different energy
density.  If $u(\br)$ is the difference between the energy densities
in the two models then we expect most of the support of $\lvert u(\br)
\rvert$ to be localized in a region similar to the shaded one in
Figure~\ref{fig:ising}.  The function $\lvert u(\br) \rvert$ should
decay exponentially to zero as you move away from the wall.  The
shaded region in Figure~\ref{fig:ising} is an \emph{energy tube} that
arises because of the boundary conditions on the wall. Other models 
of defects with non-zero correlation lengths will have the same 
property of energy localization near the defect.

A model for constructing energy tubes begins with a field
theory on $\bbE^{n}$ specified by fields $\{\varphi_{\alpha}\}$.  We
introduce an embedded submanifold $X:\Stil \to \Sigma
\hookrightarrow \bbE^{n}$ and the coupling of $\Sigma$ to the fields
is by the imposition of special boundary conditions on the submanifold
in analogy to the way a perfectly conducting plane interacts with an
electrostatic field.  Another method is $p$-brane type couplings; the
coupling of a $q$-form $A_{q}$ to $\Sigma^{q}$ is of the form
$\int_{\Sigma} A_{q}$.

From the path integral viewpoint the partition function you are 
computing is
\begin{equation}
    Z = \int [\cD X]\; e^{-T_{q} \vol(\Sigma,X) + 
    \textcolor{red}{\dotsb}} \int_{\Phi_{X}} [\cD 
    \varphi]\; e^{-I(\varphi,X)}\,,
    \label{eq:def-Z}
\end{equation}
where $X : \Sigma \hookrightarrow \bbE^{n}$ is the embedding,
$\Phi_{X}$ is the space of all field configurations compatible with
the boundary conditions imposed by the embedding of $\Sigma$, and
$I(\varphi,X)$ is the action for the fields.  The integral over
$\varphi$ depends on $\Sigma$ via the map $X$ because the coupling
of $\Sigma$ to the fields $\varphi$ is through the boundary
conditions.  The free energy due to the fields $\varphi$ is
\begin{equation*}
    e^{-F(X)} = \int_{\Phi_{X}} [\cD 
    \varphi]\; e^{-I(\varphi,X)}\,.
\end{equation*}
Let $F_{0}$ be the free energy due to $\varphi$ in the absence of 
$\Sigma$, and let $(\Delta F)(X) = F(X) - F_{0}$ be the relative free 
energy. The full partition 
function \eqref{eq:def-Z} becomes 
\begin{equation}
    Z = e^{-F_{0}}\; \int [\cD X]\; e^{-T_{q} \vol(\Sigma,X) - 
    (\Delta F)(X)}\,.
    \label{eq:Z-1}
\end{equation}
The problem is computing $\Delta F$.  We will use some mean field
theory arguments to propose a form for $\Delta F$.  A pr\'{e}cis of
the subsequent paragraphs is that if all fields $\varphi$ are massive
then the presence of $\Sigma$ leads to a relative free energy density
function\footnote{We use the letter $u$ as a generic energy density
whether it be regular energy or free energy.} $u$ that should decay
exponentially as you move away from $\Sigma$.  The energy tube is
located where $\lvert u\rvert$ is ``large''.  An energy tube does not have
a fixed radius as a Weyl tube; the energy density $\lvert u \rvert$
drops off as you move away from $\Sigma$.  We assume a  mean field type
form for the energy given by an energy density $u$:
\begin{equation}
    (\Delta F)(X) = \int_{\bbE^{n}} u(\bx,X) \; d^{n}x\,.
    \label{eq:def-delF}
\end{equation}
We emphasize in the formula above that the free energy density 
depends on the embedding $X : \Sigma \hookrightarrow \bbE^{n}$. A 
generalization of Weyl's formula for the volume of a tube will be 
used to evaluate \eqref{eq:def-delF} and show that it has a universal 
form depending of a finite number of terms.

Another well known example of boundary conditions changing the energy
density is the Casimir effect.  For the discussion here we switch to
quantum field theory language in Minkowski space $\bbM^{4}$.  The
field theory is free photons in the presence of two perfectly
conducting planes.  The electromagnetic field has to satisfy some
special boundary conditions.  The energy of the system is determined
by the zero point fluctuations of the electromagnetic field.  The
normal modes for the electromagnetic field in the system with plates
are different from the ones without plates resulting an attractive
force between the plates.  This is a system with massless particles
and therefore the discussions given in this article are not valid.

\section{Weyl's volume element formula}
\label{sec:vol-element}

Even though Weyl's paper is about the volume of tubes what he actually
derives is a formula expressing the Euclidean volume element $d^{n}x$
in terms of the volume element on the submanifold $\Sigma$, and ``polar
coordinates orthogonal to $\Sigma$''.  We provide a more modern
derivation of Weyl's result \cite{Weyl:tubes} below.  Weyl's
expression for the volume element is what allows us to consider a
multipole expansion for the energy.

\begin{figure}[tbp]
    \centering
    \includegraphics[width=0.5\textwidth]{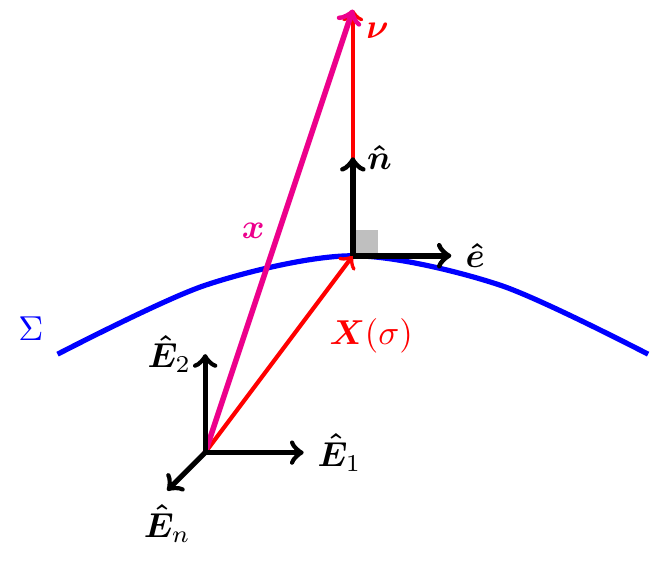}
    \caption[coords]{Representation of a point $\bx \in
    \bbE^{n}$ in terms of a point $\bX(\sigma)$ on the surface
    $\Sigma$ and a vector $\bnu$ orthogonal to the surface: $\bx =
    \bX(\sigma) + \bnu$.}
    \label{fig:coordinates}
\end{figure}
Let $(\bE_{1},\ldots, \bE_{n})$ be a standard orthonormal cartesian
basis for $\bbE^{n}$.  If $\bx \in \bbE^{n}$ then we assign
coordinates to that point by $\bx = \bE_{\mu}x^{\mu}$.  Let the map
$X: \Stil^{q} \to \bbE^{n}$ be an embedding of the image $X(\Stil)=
\Sigma \hookrightarrow \bbE^{n}$ in Euclidean space. Then for each
$\sigma \in \Sigma \subset \bbE^{n}$ we have an orthogonal direct sum
decomposition of the tangent bundle $T_{\sigma}\bbE^{n} =
T_{\sigma}\Sigma \oplus \left(T_{\sigma}\Sigma\right)^{\perp}$ where
$(T_{\sigma}\Sigma)^{\perp}$ is the normal bundle at $\sigma
\in\Sigma$.  In a small tubular neighborhood of $\Sigma$, we can
describe a point $\bx$ uniquely by
\begin{equation}
    \bx = \bX(\sigma) + \bnu\,, 
    \label{eq:def-bx}
\end{equation}
where $\bnu \in (T_{\sigma}\Sigma)^{\perp}$, see
Figure~\ref{fig:coordinates}.  Locally choose an orthonormal frame
$(\ehat_{1}, \dotsc, \ehat_{q})$ for $T\Sigma$ and an orthonormal
frame $(\nhat_{q+1}, \dotsc, \nhat_{n})$ for $(T\Sigma)^{\perp}$.
Such a frame is called a Darboux frame.  Note that $T_{\sigma}\bbE^{n}
= T_{\sigma}\Sigma \oplus (T_{\sigma}\Sigma)^{\perp}$.  Let $\iota:
\Sigma \hookrightarrow \bbE^{n}$ be the inclusion map and let
$\mathcal{F}(\bbE^{n})$ be the bundle of orthonormal frames of
$\bbE^{n}$.  The set of all Darboux frames of $\Sigma$ is a sub-bundle
of the pullback bundle $\iota^{*}\mathcal{F}(\bbE^{n}) \to \Sigma$
with structure group $\SO(q)\times \SO(l) \subset \SO(n)$, where
$n=q+l$.  Let $(\sigma^{1}, \dotsc, \sigma^{q})$ be local coordinates
on $\Sigma$ and let $\bnu = \nu^{i} \nhat_{i}$, then
$(\sigma^{1},\dotsc,\sigma^{q}, \nu^{q+1}, \dotsc, \nu^{n})$ are local
coordinates for the tubular neighborhood.  Our index convention is
that latin indices from the beginning of the alphabet go from $1$ to
$q$, latin indices from the middle of the alphabet go from $q+1$ to
$n$, and greek indices go from $1$ to $n$.  Next, we take the
differential of \eqref{eq:def-bx} and follow standard moving frame
techniques developed by E.~Cartan \cite{Cartan:riemann} (see \cite{BCG3}
for the general theory) to study submanifolds of Euclidean space.
First, note that $d\bX = \ehat_{a} \theta^{a}$ where the $\theta^{a}$
are $1$-forms on $\Sigma$ because $X:\Stil \to \bbE^{n}$, and the
``displacement of $\bX$'', $d\bX$, is tangential to the surface.  We
observe that orthonormality requires
\begin{subequations}\label{eq:def-dbasis}
\begin{align}
    d\ehat_{a} &= \ehat_{b}\; \omega_{ba} - \nhat_{j}\; K_{abj}\; 
    \theta^{b}
    \label{eq:def-de} \\
    d\nhat_{i} &= \nhat_{j}\; \omega_{ji} +
    \ehat_{a}\;K_{abi}\;\theta^{b}\;,
    \label{eq:def-dn}
\end{align}
\end{subequations}
Since $\ehat_{a}$ and $\nhat_{i}$ are defined only on $\Sigma$, it
follows that $d\ehat_{a}$ and $d\nhat_{i}$ are $1$-forms on $\Sigma$.
The $\SO(q)$ connection $1$-form on $T\Sigma$ is $\Gamma_{cab}
\theta^{c} = \omega_{ab}= -\omega_{ba}$.  The $\SO(l)$ connection
$1$-form on the normal bundle is $\Gamma_{aij}\theta^{a} =\omega_{ij}=
- \omega_{ji}$, and $K_{abi}= K_{bai}$ are the extrinsic curvatures
(second fundamental forms).  Next
we note\footnote{We are in $\bbE^{n}$ so the covariant exterior
differential is the ordinary one.} $d\bnu = (d\nhat_{i}) \nu^{i} +
\nhat_{i}\; d\nu^{i}$, and putting all this together we see that
\begin{equation}
    d\bx = \bE_{\mu}\; dx^{\mu} = \ehat_{a} \left( \delta_{ab} + 
    \nu^{i} K_{abi} \right) \theta^{b} + \nhat_{i}\; D\nu^{i}\,,
    \label{eq:dbX}
\end{equation}
where $D\nu^{i} = d\nu^{i} + \omega_{ij} \nu^{j}$ is the covariant
differential on the normal bundle.  Using the fact that $(\ehat,
\nhat)$ is an orthonormal frame and that $T_{\sigma}\Sigma \perp
(T_{\sigma}\Sigma)^{\perp}$, we see that an orthonormal coframe at
$\bx$ is given by $(\thhat^{a}, D\nu^{i}) = \bigl( (\delta_{ab}
+ \nu^{i}K_{abi})\theta^{b}, D\nu^{i}\bigr)$. Therefore, the Euclidean
metric is given by
\begin{equation}
    ds^{2}= d\bx\bdot d\bx = \delta_{ab}\thhat^{a}\otimes
    \thhat^{b} + \delta_{ij}\, D\nu^{i}\otimes D\nu^{j} =
    h_{ab}(\sigma,\nu)\; \theta^{a}\otimes \theta^{b} + \delta_{ij}\,
    D\nu^{i}\otimes D\nu^{j}\,,
    \label{eq:metric}
\end{equation}
where 
\begin{equation}
    h_{ab}(\sigma,\nu) = \bigl(I +
    \bnu\bdot\bK(\sigma)\bigr)^{2}{}_{ab}.  
    \label{eq:metric-coords}
\end{equation}

We mention that the results above are also valid in Minkowski space
$\bbM^{n}$ if the normal bundle $(T\Sigma)^{\perp}$ has Euclidean
signature, and the tangent bundle $T\Sigma$ has Lorentzian signature.
Observe that in Minkowski space $h_{ab}$ has signature
$(-,+,\dotsc,+)$ and therefore $h_{ab}$ in
eq.~\eqref{eq:metric-coords} is implicitly interpreted to be $h_{ab} =
h_{cd}^{(\mathbb{M})} (I + \bnu\bdot\bK)^{c}{}_{a} (I +
\bnu\bdot\bK)^{d}{}_{b}$ where $h_{cd}^{(\mathbb{M})}$ is the Minkowski metric
on $\bbM^{q}$.

The volume element is easily computed in the orthonormal coframe
\begin{align*}
    d^{n}x &= \thhat^{1}\wedge\thhat^{2}\wedge \dotsb
    \wedge \thhat^{q}\wedge D\nu^{q+1} \wedge \dotsb \wedge
    D\nu^{n}\,,\\
    &= \det(I + \bnu\bdot\bK)\;
    \theta^{1}\wedge \theta^{2} \wedge \dotsb \wedge \theta^{q} \wedge
    D\nu^{q+1} \wedge \dotsb \wedge D\nu^{n}\,.
\end{align*}
The maximum number of powers of $\theta$ is already present in the
expression above, hence we can replace $D\nu$ by $d\nu$  and
obtain Weyl's formula for the Euclidean volume element in a tubular
neighborhood of $\Sigma$:
\begin{align}
    d^{n}x &= \det(I + \bnu\bdot\bK)\; \theta^{1}\wedge \theta^{2} \wedge
    \dotsb \wedge \theta^{q} \wedge d\nu^{q+1} \wedge d\nu^{q+2}
    \wedge \dotsb \wedge d\nu^{n}\,,
    \nonumber \\
    &= \det(I + \bnu\bdot\bK)\; \dual_{\Sigma} \wedge
    d\nu^{q+1} \wedge d\nu^{q+2} \wedge \dotsb \wedge d\nu^{n}\,,
    \label{eq:Weyl-vol}
\end{align}
where $\dual_{\Sigma}$ is the volume element on $\Sigma$.  

We mention that the Weyl volume element formula \eqref{eq:Weyl-vol} is
also valid in Minkowski space $\bbM^{n}$ as long as the submanifold
$\Sigma$ is timelike.  Note that the normal tangent bundle
$(T_{\sigma}\Sigma)^{\perp}$ has Euclidean signature and our
derivations are valid.  The volume element $\dual_{\Sigma}$ is now the
Lorentzian volume element on $\Sigma$.

Taking a Minkowski space field theory viewpoint, we point out that
$\sqrt{-\det g}\; g^{\mu\nu}$ enters into the Lagrangian for scalar
fields and $\sqrt{-\det g}\; g^{\mu\nu}g^{\kappa\lambda}$ enters into
the Lagrangian for vector fields, and these can be computed in the
Darboux frame using the formula \eqref{eq:metric} for the metric.  In
computing the action of a field theory we expect to find
terms of the type
\begin{subequations}\label{eq:energy-qft}
    \begin{align}
	E & = \int_{\Sigma} \dual_{\Sigma}(\sigma) \left(
	\int_{(T_{\sigma}\Sigma)^{\perp}} d^{l}\nu \;\det(I +
	\bnu\bdot\bK)\; u(\sigma, \bnu)\right) 
	\label{eq:energy-0}\\
	&\quad +  \int_{\Sigma} \dual_{\Sigma}(\sigma) \left(
	\int_{(T_{\sigma}\Sigma)^{\perp}} d^{l}\nu \;\det(I +
	\bnu\bdot\bK)\; h^{ab}(\sigma,\bnu)\; v_{ab}(\sigma, \bnu) \right)
	\label{eq:energy-2} \\
	&\quad +  \int_{\Sigma} \dual_{\Sigma}(\sigma) \left(
	\int_{(T_{\sigma}\Sigma)^{\perp}} d^{l}\nu \;\det(I +
	\bnu\bdot\bK)\; h^{ab}(\sigma,\bnu)\, h^{cd}(\sigma,\bnu)\;
	w_{abcd}(\sigma,\bnu)\right).
	\label{eq:energy-4}
    \end{align}
\end{subequations}
The main result of this paper is a general formula for the potential
energy like term \eqref{eq:energy-0} which involves a finite number of
monopole moments that is a generalization of Weyl's volume formula.
The kinetic energy like terms, \eqref{eq:energy-2} and
\eqref{eq:energy-4}, lead to an infinite power series in $K_{ab}{}^{j}$.
We have not found a re-expression of those formulas that leads to
anything resembling the simplicity of Weyl's volume formula.  As a
field theory on $\Sigma$, we note that time derivatives will only
appear in \eqref{eq:energy-2} and \eqref{eq:energy-4}.  For the rest
of this paper we concentrate on evaluating \eqref{eq:energy-0}.

\section{The energy contained in an energy tube}
\label{sec:energy}

Let $u(\bx)$ be the energy density at $\bx\in\bbE^{n}$, then the
energy in a small volume $d^{n}x$ is given by $u(\bx)\; d^{n}x$.
Because we are interested in the energy density near $\Sigma$, it is
convenient to use Weyl's volume element \eqref{eq:Weyl-vol} formula to
express the total energy as
\begin{equation}
    E = \int_{\Sigma} \dual_{\Sigma}(\sigma) \left( 
    \int_{(T_{\sigma}\Sigma)^{\perp}} u(\sigma, \bnu)\; \det(I + \bnu\bdot\bK)\;
    d^{l}\nu \right).
    \label{eq:total-energy}
\end{equation}
In general, the normal bundle integral cannot be performed.  Here we
show that if $u(\sigma,\bnu)$ decays rapidly enough away from $\Sigma$
then it has a multipole expansion, and the normal bundle integral can
be simplified and expressed in terms of a \emph{finite} number of
radial moments associated with the energy density.  The derivation of
the general formula is quite technical and we defer it for the moment;
instead, we first discuss what the answer looks like in the important
special case of spherical symmetry.

\subsection{The spherically symmetric case}
\label{sec:spher-sym}

There are interesting examples where the energy density function
$u(\sigma,\bnu)$ is spherically symmetric in the normal bundle,
namely, $u(\sigma,\bnu) = u^{(0)}(\sigma,\lVert \bnu \rVert)$.  We
refer to this as a monopole energy density.  The energy is given by
\begin{equation}
    E^{(0)} = \int_{\Sigma} \dual_{\Sigma}\int_{(T_{\sigma}\Sigma)^{\perp}}
	u^{(0)}(\sigma, \lVert\bnu\rVert)\; \det(I +
	\bnu\bdot\bK)\; d^{l}\nu \,.
    \label{eq:E-monopole}
\end{equation}
First we define some curvature invariants associated with the
intrinsic geometry of $\Sigma$.  The curvature $2$-form is
$\Omega_{ab} = \half R_{abcd}\theta^{c}\wedge\theta^{d}$, the Hodge
dual of $\theta^{a_{1}}\wedge \cdots\wedge\theta^{a_{2r}}$ is
$\dual^{a_{1}a_{2}\dotsm a_{2r-1}a_{2r}}$, and other notations are
explained in Appendix~\ref{sec:misc}.  Let
\begin{equation}
    \begin{split}
    \curv_{2r}(\Sigma)\; \dual_{\Sigma} &= \frac{1}{4^{r}\, r!}\; 
    \delta_{a_{1}\dotsm a_{2r}}^{b_{1}\dotsm
    b_{2r}}\; R_{a_{1}a_{2}b_{1}b_{2}} \dotsm
    R_{a_{2r-1}a_{2r}b_{2r-1}b_{2r}}\, \dual_{\Sigma}\,, \\
    &= \frac{1}{2^{r}\, r!}\; \dual^{a_{1}a_{2}\dotsm a_{2r-1}a_{2r}}
    \wedge \Omega_{a_{1}a_{2}} \wedge\dotsb \wedge
    \Omega_{a_{2r-1}a_{2r}}\,.
    \end{split}
    \label{eq:def-curv-x}
\end{equation}
The first three $\curv_{2r}$ are:
\begin{equation}
    \begin{split}
	\curv_{0}(\Sigma) &= 1\,, \\
	\curv_{2}(\Sigma) & = \frac{1}{2}\; R\,, \\
	\curv_{4}(\Sigma) &= \frac{1}{8} \left(R^{2} - 4 R_{ab}R^{ab}
	+ R_{abcd}R^{abcd}\right).
    \end{split}
    \label{eq:K-table-x}
\end{equation}
The $\int_{\Sigma}\curv_{2r}\, \dual_{\Sigma}$ defined here are the
same as the $k_{2r}(\Sigma)$ in Gray~\cite[p.~56]{Gray:tubes}. These 
are the curvature combinations identified by Lovelock that lead to 
generalizations of the Einstein tensor.

If $\dim \Sigma =q=2r$ is even then the differential form  of
maximal degree is
\begin{equation*}
    \curv_{q}(\Sigma) \; \dual_{\Sigma} =
    \frac{1}{2^{r}\, r!}\; \epsilon^{a_{1}a_{2}\dotsm
       a_{2r-1}a_{2r}} \; \Omega_{a_{1}a_{2}} \wedge\dotsb \wedge
       \Omega_{a_{2r-1}a_{2r}} = \pf(\Omega) \; \dual_{\Sigma}\,,
\end{equation*}
where $\pf(\Omega) = \curv_{q}(\Sigma)$ is the Pfaffian of the
``antisymmetric matrix valued $2$-form'' $\Omega_{ab}$.  The Euler
characteristic of $\Sigma$ is $\chi(\Sigma) = (1/2\pi)^{q/2}
\int_{\Sigma}\pf(\Omega)\,\dual_{\Sigma}$ by the general Gauss-Bonnet
Theorem of Chern.

Next we define the ``normal radial moments'' of the monopole energy
density by
\begin{equation}
    \mu^{(0)}_{2r}(\sigma) = \int_{(T_{\sigma}\Sigma)^{\perp}} 
    \lVert\bnu\rVert^{2r}\; 
    u^{(0)}(\sigma,\lVert\bnu\rVert)\; d^{l}\nu = V_{l-1}(S^{l-1}) \int_{0}^{\infty} 
    d\nu\; \nu^{2r+l-1}\; u^{(0)}(\sigma,\nu)\;,
    \label{eq:def-mu2r-x}
\end{equation}
where $V_{l-1}(S^{l-1})$ is the $(l-1)$-volume of $S^{l-1}$, see
Appendix~\ref{sec:avg-sphere}.  The energy of the tube is given by
\begin{equation}
    E^{(0)} = \sum_{r=0}^{\lfloor q/2 \rfloor} C_{2r}\; \int_{\Sigma} 
    \mu^{(0)}_{2r}(\sigma)\; \curv_{2r}(\Sigma)\; \dual_{\Sigma}\,,
    \label{eq:gen-Weyl-x}
\end{equation}
where $C_{2r}$ are constants specified by \eqref{eq:def-C}.  Note that
the energy only depends on the intrinsic geometry of $\Sigma$.  The
derivation of this formula is presented in detail in
Section~\ref{sec:monopole}.  Is equation \eqref{eq:gen-Weyl-x}
actually an equality or is it an approximation?  We defer this
question to a post-derivation discussion in
Section~\ref{sec:monopole}.

The discussion of Section~\ref{sec:physics-tubes} can now be applied
to the monopole contribution to the energy \eqref{eq:gen-Weyl-x}.  The
effect of the bulk QFT that lives on $\bbE^{n}$ is to produce
effective scalar fields $\mu^{(0)}_{0}(\sigma), \mu^{(0)}_{2}(\sigma),
\mu^{(0)}_{4}(\sigma), \dotsc, \mu^{(0)}_{2\lfloor
q/2\rfloor}(\sigma)$ that couple linearly to Lovelock type curvature
terms.  Note that these effective fields live on $\Sigma$ and in the
very small curvature situation we can approximate them as being
constants.  To ``leading order'' we would replace the fields
$\mu^{(0)}_{2r}(\sigma)$ by constants $\bar{\mu}^{(0)}_{2r}$.  The net
effect of the bulk QFT to leading order is to induce an effective
action for the surface that is a Lovelock type gravitational action.

The symmetric Lovelock tensors $\love{2r}^{ab}$ are the
generalizations of the Einstein tensor proposed by Lovelock.  The
$2r$-th one is defined by the variational derivative of the parent
Lovelock action $I_{2r}$, see~\eqref{eq:Lovelock-lag-0} or 
\eqref{eq:gen-Weyl-x},
\begin{equation}
    \delta I_{2r} = \delta \int_{\Sigma} \curv_{2r}\, \dual_{\Sigma}
    = -\half \int_{\Sigma} \love{2r}^{ab}\;(\delta g_{ab}) 
    \dual_{\Sigma}
    \label{eq:love-tensor}
\end{equation}
Since each action $I_{2r}$ is diffeomorphism invariant, the
corresponding Lovelock tensor is automatically conserved
$D_{a}\love{2r}^{ab}=0$.  For example, the Lovelock tensor $E_{2}$ is
precisely the Einstein tensor $E_{2}^{ab} = R^{ab} - \half 
g^{ab}\,R$. A straightforward computation gives
\begin{equation}
    \love{2r}{}^{c}{}_{b} = -\frac{1}{4^{r}\, r!}\; 
     \delta^{cd_{1}d_{2}\dotsm d_{2r-1}d_{2r}}_{ba_{1}a_{2}\dotsm 
     a_{2r-1}a_{2r}}\; 
     R^{a_{1}a_{2}}{}_{d_{1}d_{2}}\dotsm 
     R^{a_{2r-1}a_{2r}}{}_{d_{2r-1}d_{2r}}\,.
    \label{eq:def-love-2r}
\end{equation}
If $q$ is even then $\love{q}^{ab}=0$ because the multi-index
$\delta$-tensor is identically zero since it involves $q+1$ total
antisymmetrizations.  The vanishing of $\love{q}$ is a consequence of
the topological nature of $\curv_{q}$.

\subsection{The volume of a tube}
\label{sec:Weyl-vol}

An example of the spherically symmetric case is an energy tube with
constant energy density $u_{0}$ out to a radius $\rho$ and vanishing 
beyond:
\begin{equation}
    u(\sigma,\bnu) = u^{(0)}(\sigma,\nu) = 
    \begin{cases}
	u_{0} & \nu \le \rho\,, \\
	0 & \nu > \rho\,.
    \end{cases}
    \label{eq:u-constant}
\end{equation}
The moments are given by
\begin{equation}
    \mu^{(0)}_{2r} = u_{0}\, V_{l-1}(S^{l-1})\; \frac{\rho^{l+2r}}{l+2r}\;.
    \label{eq:moments-constant}
\end{equation}
In this case the energy will be $E = u_{0} \vol(\tube(\Sigma,\rho))$ 
and we obtain Weyl's formula for the volume of a tube:
\begin{equation*}
    \vol(\tube(\Sigma,\rho)) = V_{l-1}(S^{l-1}) \rho^{l}
    \sum_{r=0}^{\lfloor q/2 \rfloor} \frac{C_{2r}}{l+2r}\; \rho^{2r}
    \int_{\Sigma} \frac{1}{2^{r}\, r!}\; \dual^{a_{1}a_{2}\dotsm
    a_{2r-1}a_{2r}} \wedge \Omega_{a_{1}a_{2}} \wedge\dotsb \wedge
    \Omega_{a_{2r-1}a_{2r}}.
\end{equation*}
Note that
\begin{equation*}
    \frac{C_{2r}}{l+2r} = \prod_{k=0}^{r} \frac{1}{l+2k}\;.
\end{equation*}
Using \eqref{eq:vol-ball} and putting it all together, we have Weyl's
formula for the volume of a tube:
\begin{equation}
    \vol(\tube(\Sigma,\rho)) = V_{l}(B^{l}) \rho^{l}
    \sum_{r=0}^{\lfloor q/2 \rfloor}
    \frac{l\, \rho^{2r}}{\prod_{k=0}^{r} (l+2k)} 
    \int_{\Sigma} \curv_{2r}(\Sigma)\; \dual_{\Sigma}\;.
    \label{eq:Weyl-tube-for}
\end{equation}
The $\int_{\Sigma}\curv_{2r}\, \dual_{\Sigma}$ defined here are the
same as the $k_{2r}(\Sigma)$ in Gray~\cite[p.~56]{Gray:tubes}.  The equation
above may also be written as
\begin{equation}
    \vol_{n}(\tube(\Sigma,\rho)) = V_{l}(B^{l}) \rho^{l} \;
    \vol_{q}(\Sigma)
    + V_{l}(B^{l}) 
    \sum_{r=1}^{\lfloor q/2 \rfloor}
    \frac{\rho^{l+2r}}{\prod_{k=1}^{r} (l+2k)} 
    \int_{\Sigma} \curv_{2r}(\Sigma)\; \dual_{\Sigma}\;.
    \label{eq:Weyl-tube-for-1}
\end{equation}

\subsection{Dirac-Born-Infeld action}
\label{sec:DBI}

The effective action for D-branes is the Dirac-Born-Infeld action, for
a review see~\cite{Johnson:D-branes}.  The DBI action in a coordinate
basis is usually written as $\int_{\Sigma} \sqrt{\det(g + B)}\;
d^{q}\sigma$ where $B$ is an antisymmetric $2$-tensor.  If we go to an
orthonormal frame we see that the DBI action is
\begin{equation*}
    \int_{\Sigma}\sqrt{\det\left(I + \widetilde{B}\right)}\; \dual_{\Sigma}
\end{equation*}
where the matrix elements of $\widetilde{B}$ are the components of the
antisymmetric $2$-form in the orthonormal frame.  This corresponds to
a monopole moment of the form $\mu_{0}(\sigma) \propto
\sqrt{\det\left(I + \widetilde{B}\right)}$.

\section{The multipole expansion for the energy}
\label{sec:multipole}

\subsection{Spherical harmonics}
\label{sec:harmonics}

\begin{figure}[tbp]
    \centering
    \includegraphics[width=0.4\textwidth]{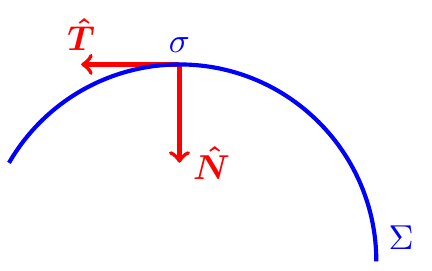}
   \caption[xxx6u8]{In general, the  energy density
   $u(\sigma,\bnu)$ will not be isotropic.  Consider the example
   of a planar curve $\Sigma \hookrightarrow \bbE^{3}$.  Here
   $\bm{\hat{T}}$ is the unit tangent vector, $\bm{\hat{N}}$ is
   the unit normal and $\bm{\hat{B}} = \bm{\hat{T} \times
   \hat{N}}$ is the binormal in the standard Frenet-Serret framing.  Note
   that $(T_{\sigma}\Sigma)^{\perp}$ is spanned by the normal and
   the binormal but we do not expect the energy density to be
   isotropic in $(T_{\sigma}\Sigma)^{\perp}$ because of the large
   curvature.}
    \label{fig:tube-non-iso}
\end{figure}

In general, the energy density $u(\sigma,\bnu)$ will not be spherically
symmetric, see Figure~\ref{fig:tube-non-iso}.  To proceed with the
evaluation of \eqref{eq:total-energy} we use a multipole expansion.
Let $\lVert \bnu \rVert = \nu$ and  write $\bnu = \nu\, \bnuhat$
where $\lVert \bnuhat \rVert=1$.
Group representation theory tells us that the real valued functions on
$S^{l-1}$ decompose into an orthogonal direct sum
$\bigoplus_{j=0}^{\infty} W^{j}$ of real finite dimensional
irreducible representation spaces $W^{j}$ of $\SO(l)$, \emph{i.e.},
$W^{j}$ is an irreducible real $\SO(l)$-module.  The standard
expansion for $u(\sigma,\bnu)$ in terms of \emph{spherical harmonics}
is given by
\begin{equation}
    u(\sigma,\bnu) = \sum_{j=0}^{\infty}\sum_{M=1}^{\dim W^{j}} 
    u^{(j)}_{M}(\sigma,\lVert \bnu \rVert)\; Y^{j}{}_{M}(\bnuhat)\,,
    \label{eq:def-multipole}
\end{equation}
where $\{ Y^{j}{}_{M}(\bnuhat) \}$ is a real orthogonal
basis of spherical harmonics for the sphere $S^{l-1}$ with
normalization
\begin{equation}
    \int_{S^{l-1}} Y^{j}{}_{M}(\bnuhat) Y^{j'}{}_{M'}(\bnuhat)\;
    d\vol_{S^{l-1}} = \frac{V(S^{l-1})}{\dim W^{j}}\;
    \delta_{jj'}\delta_{MM'}\,.
    \label{eq:Y-norm}
\end{equation}
The index $j$ refers to the ``spin $j$'' representation of $\SO(l)$; 
we will make this  precise soon.
A result we wish to establish in this section is that the total
energy \eqref{eq:total-energy} only depends on the first $q$ multipole
moments $u^{(j)}{}_{M}$ of \eqref{eq:def-multipole}, \emph{i.e.}, $0 \le
j \le q$.  To prove this observation and to discuss the evaluation of
the integral \eqref{eq:total-energy}, it is better to use
\emph{cartesian spherical harmonics} that we will describe shortly.
The proof requires understanding the relationship between polynomials
in the variables $\nuhat^{1},\dotsc, \nuhat^{l}$ and the spherical
harmonics.

Think of the cartesian coordinates $(\nu^{1},\nu^{2}, \dotsc,
\nu^{l})$ as a basis for the real vector space\footnote{A coordinate
function on vector space is a linear functional and therefore the
coordinate functions are basis for the dual vector space of $\bbE^{l}$
but because we have a metric we implicitly identify $W \approx
\bbE^{l}$ with its dual space $W^{*}$.} $W \approx \bbE^{l}$.  In
other words, we write $v\in W$ as a linear functional $v =
c_{i}\nu^{i}$.  $\SO(l)$ acts on $W$ irreducibly.  The set of
homogenous polynomials of degree $k$ are isomorphic to the $k$-fold
symmetric tensor product $\Sym^{k}W$.  The space $\Sym^{k}W$ is
invariant under the action of $\SO(l)$ but the action is not
irreducible because of the existence of the trace operation.  The
monomials $\nu^{i_{1}}\dotsm \nu^{i_{k}}$ span $\Sym^{k}W$ but tracing
on the last two indices leads to monomials $\nu^{i_{1}}\dotsm
\nu^{i_{k-2}} \lVert \bnu \rVert^{2}$ and the span of these
homogeneous polynomials of degree $k$ (isomorphic to $\lVert \bnu
\rVert^{2} \cdot \Sym^{k-2}W$) is an invariant subspace of
$\Sym^{k}W$.  Our irreducible representation space $W^{j}$ is
isomorphic to the vector space of rank $j$ symmetric traceless
tensors.  A spanning set related to the spherical harmonics on the
sphere may be constructed from the homogeneous degree $j$ polynomials
in the following way.  We restrict the homogeneous polynomials to the
unit sphere by imposing the condition $\lVert \bnu \rVert^{2}=1$.
Restriction to the sphere turns a homogeneous degree $j$ polynomial
into an inhomogeneous polynomial with respect to degree because every
occurrence of $\sum_{i} \nuhat^{i}\nuhat^{i}$ is replaced by $1$.  We
will refer to the spanning set as the \emph{faux} cartesian spherical
harmonics on $S^{l-1}$ because they are not a basis but an \emph{over
complete} spanning set.  Our notation for the faux spherical harmonics
is $\Yh^{j}_{i_{1}i_{2} \dotsm i_{j}}(\bnuhat)$.  The first few faux
cartesian spherical harmonics are
\begin{equation}
    \begin{split}
	\Yh^{0}(\bnuhat) &= 1 \,,\\
	\Yh^{1}_{i}(\bnuhat) &= \nuhat^{i}\,,\\
	\Yh^{2}_{ii'}(\bnuhat) &= 
	\nuhat^{i}\nuhat^{i'} - 
	\frac{1}{l}\; \delta^{ii'} \,,
	\\
	\Yh^{3}_{i_{1}i_{2}i_{3}}(\bnuhat) &= 
	\nuhat^{i_{1}}\nuhat^{i_{2}}\nuhat^{i_{3}} -\frac{1}{l+2} 
	\left[\delta^{i_{1}i_{2}}\nuhat^{i_{3}} + 
	\delta^{i_{2}i_{3}}\nuhat^{i_{1}} + 
	\delta^{i_{3}i_{1}}\nuhat^{i_{2}} \right] \;.
    \end{split}
    \label{eq:Y-first}
\end{equation}
For $l>1$, the faux cartesian spherical harmonics are uniquely
specified by
\begin{enumerate}
    \item $\Yh^{j}_{i_{1}i_{2} \dotsm i_{j}}(\bnuhat)$ is totally
    symmetric under any permutation of $i_{1},i_{2},\dotsc, i_{j}$.
    
    \item $\Yh^{j}_{i_{1}i_{2} \dotsm i_{j}}(\bnuhat)$ is traceless
    with respect to contraction on any pair of indices.  Because the
    faux harmonic is totally symmetric in the lower indices, this
    reduces to $\Yh^{j}_{iii_{3} \dotsm i_{j}}(\bnuhat)=0$.
    
    \item The parity of $\Yh^{j}$ is $(-1)^{j}$.
    
    \item $\Yh^{j}_{i_{1}i_{2} \dotsm i_{j}}(\bnuhat)$ is an
    inhomogeneous polynomial of degree $j$ in the $\nuhat^{i}$ with
    normalization determined by
    \begin{equation*}
	\Yh^{j}_{i_{1}i_{2} \dotsm i_{j}}(\bnuhat) = 
	\nuhat^{i_{1}}\nuhat^{i_{2}} \dotsm \nuhat^{i_{j}} +
	(\text{polynomial of degree $j-2$}) 
	\,.
    \end{equation*}
\end{enumerate}

Let $W$ be the defining representation space of $\SO(l)$ and let
$\Sym^{k}W$ denote the $k$-th symmetric tensor product, $\dim
\Sym^{k}W = \tbinom{l+k-1}{k} = l(l+1) \dotsm (l+k-1)/k!$.  For fixed
$j$, the span of $\{\Yh^{j}_{i_{1}i_{2} \dotsm i_{j}}(\bnuhat)\}$ is a
real irreducible $\SO(l)$-module (an irreducible representation in a
real vector space) that we denote by $W^{j}$.  The representation
space $W^{j}$ is isomorphic to the space of symmetric traceless
tensors of rank $j$.  The $\{\Yh^{j}_{i_{1}i_{2} \dotsm
i_{j}}(\bnuhat)\}$ are in general not linearly independent.  For
example, if $l=3$ then the tracelessness condition tells us that
$\Yh^{2}_{11} = - \Yh^{2}_{22}- \Yh^{2}_{33}$.  The spherical
harmonics $\{Y^{j}_{M}\}$ are a basis for $W^{j}$ but the faux
harmonics $\{\Yh^{j}_{i_{1}i_{2} \dotsm i_{j}}\}$ are an over complete
spanning set if $j\ge 2$.  For $j=0$ and $j=1$ we can choose $Y^{0}
\propto \Yh^{0}$ and $Y^{1}{}_{i} \propto \Yh^{1}_{i}$.

Next we define \emph{cartesian multipole moments} by writing
\begin{equation}
    u(\sigma,\bnu) = \sum_{j=0}^{\infty}\sum_{i_{1},\dotsc, i_{j}} 
    u^{(j)}_{i_{1}\dotsm i_{j}}(\sigma, \lVert \bnu \rVert) \;
    \Yh^{j}_{i_{1} \dotsm i_{j}}(\bnuhat)\,.
    \label{eq:def-multipole-c}
\end{equation}
In the above we require $u^{(j)}_{i_{1}\dotsm i_{j}}$ to be totally
symmetric and traceless in the indices $i_{1}\dotsm i_{j}$.  The
symmetric traceless tensor $u^{(j)}_{i_{1}\dotsm i_{j}} \;
D\nu^{i_{1}} \otimes \dotsb \otimes D\nu^{i_{j}}$ is well defined.

$\Sym^{k}W$ is a reducible representation and we have a direct sum
decomposition into irreducible representations \cite[Exercise
19.21]{FultonHarris}
\begin{equation}
    \Sym^{k}W = \bigoplus_{r=0}^{\lfloor k/2 \rfloor} W^{k-2r}\,.
    \label{eq:CG-series}
\end{equation}
You can verify that $\dim W^{k} = \dim \Sym^{k}W - \dim \Sym^{k-2}W$, $k \ge
2$, and 
\begin{equation}
    \dim W^{k} = \frac{(l-1)(l+2k-2)}{k(k-1)} 
    \binom{l+k-3}{k-2}\;,\quad k\ge 2.
    \label{eq:dim-Wk}
\end{equation}

To proceed with the evaluation of the energy we have to make some
assumptions about $u(\sigma,\bnu)$.  Not all these assumptions are
required by the mathematics, but they are motivated by physics
considerations.  We require $u(\sigma,\bnu)$ to decay rapidly enough as
$\lVert \bnu\rVert \to \infty$.  In many examples you have exponential
decay.  We are interested in weak gravity and thus we expect the
curvature of $\Sigma$ to be small.  To first approximation $\Sigma
\approx \bbE^{q}$ and in this case we expect the energy density to be
translationally invariant with respect to $\Sigma$ and probably
spherically symmetric in the normal bundle direction.  As the surface
starts to curve we expect that the higher multipole moments are built
up.  Schematically we assume a hierarchy with $u^{(0)}(\sigma,\nu)
\succ u^{(1)}(\sigma,\nu) \succ \dotsb \succ u^{(q)}(\sigma,\nu)$
where we use the symbol $\succ$ to indicate a vague hierarchical
structure in magnitude.  We also assume the $u^{(j)}(\sigma,\nu)$ are
slowly varying functions of $\sigma$.

To evaluate the energy we focus on the parenthetical expression in
\eqref{eq:total-energy}, apply \eqref{eq:det-expansion} and obtain
\begin{equation}
    \begin{split}
	& \int_{\Sigma} \dual_{\Sigma}\int_{(T_{\sigma}\Sigma)^{\perp}}
	u(\sigma, \bnu)\; \det(I +
	\bnu\bdot\bK)\; d^{l}\nu \\
	&= \int_{\Sigma} \dual_{\Sigma}\sum_{r=0}^{q}
	\int_{\nu=0}^{\infty}\int_{S^{l-1}} u(\sigma, \bnu)\;
	\frac{\nu^{r}}{r!}\; \delta_{a_{1}\dotsm a_{r}}^{b_{1}\dotsm
	b_{r}}\; K^{a_{1}}{}_{b_{1}i_{1}}(\sigma)
	K^{a_{2}}{}_{b_{2}i_{2}}(\sigma) \dotsm
	K^{a_{r}}{}_{b_{r}i_{r}}(\sigma)\\
	&\qquad\qquad \times
	\nuhat^{i_{1}}\nuhat^{i_{2}}\dotsm \nuhat^{i_{r}} \cdot
	\nu^{l-1}\; d\nu\; d\vol_{S^{l-1}}\;.
    \end{split}
    \label{eq:W-int}
\end{equation}
In the expression above we have to perform the angular integral
\begin{equation}
    \int_{S^{l-1}} u(\sigma,\bnu)\; \nuhat^{i_{1}}\nuhat^{i_{2}}\dotsm
    \nuhat^{i_{r}} \;d\vol_{S^{l-1}} \quad\text{for $0 \le r \le q$.}
    \label{eq:ang-int}
\end{equation}
To understand the result of this integration we use the decomposition
\eqref{eq:CG-series} and expand using our basis of spherical
harmonics:
\begin{equation*}
	\nuhat^{i_{1}}\nuhat^{i_{2}}\dotsm \nuhat^{i_{r}} = 
	\sum_{s=0}^{\lfloor r/2 \rfloor} \; \sum_{M=1}^{\dim W^{r-2s}}
	B^{i_{1}\dotsm i_{r}}_{r-2s,M}\;
	Y^{r-2s}{}_{M}(\bnuhat)\,,
\end{equation*}
for some constant coefficients $B^{\bullet}_{\bullet}$.  Note that in
the sum above you only get spherical harmonics with parity $(-1)^{r}$.
Inserting this expansion into the angular integration
\eqref{eq:ang-int} we see that result of the integration can only
involve the multipoles $u^{(r)}_{\bullet}, u^{(r-2)}_{\bullet}, \dots,
u^{(r-2\lfloor r/2\rfloor)}_{\bullet}$.  Next we observe that in the
basic integral \eqref{eq:W-int}, the sum over $r$ goes from $0$ to $q$
and therefore the energy of the energy tube will only depend on the
first $q$ multipole moments
$u^{(0)}_{\bullet},u^{(1)}_{\bullet},\dotsc, u^{(q-1)}_{\bullet},
u^{(q)}_{\bullet}$.  This result is surprising.  Higher multipole
moments for the energy density correspond to finer grained angular
resolution in the normal tangent space $(T_{\sigma}\Sigma)^{\perp}$.
The total energy of an energy tube is insensitive to variations in the
energy density $u(\sigma,\bnu)$ on angular scales smaller than roughly
$2\pi/q$ radians.

\subsection{The monopole contribution}
\label{sec:monopole}

To evaluate the monopole contribution to the energy we return to
equation \eqref{eq:W-int} and replace $u(\sigma,\bnu)$ by
$u^{(0)}(\sigma,\lVert \bnu \rVert)$ and obtain
\begin{align*}
	& \int_{\Sigma} \dual_{\Sigma}\int_{(T_{\sigma}\Sigma)^{\perp}}
	u(\sigma, \bnu)\; \det(I +
	\bnu\bdot\bK)\; d^{l}\nu \\
	&= \int_{\Sigma} \dual_{\Sigma}\sum_{r=0}^{q}
	\int_{\nu=0}^{\infty}\int_{S^{l-1}} u^{(0)}(\sigma, \nu)\; 
	\frac{\nu^{r}}{r!}\; \delta_{a_{1}\dotsm a_{r}}^{b_{1}\dotsm 
	b_{r}}\; K^{a_{1}}{}_{b_{1}i_{1}}K^{a_{2}}{}_{b_{2}i_{2}} 
	\dotsm K^{a_{r}}{}_{b_{r}i_{r}}\\
	&\qquad\qquad \times
	\nuhat^{i_{1}}\nuhat^{i_{2}}\dotsm \nuhat^{i_{r}} \cdot
	\nu^{l-1}\; d\nu\; d\vol_{S^{l-1}}\;,\\
	&= \int_{\Sigma} \dual_{\Sigma} \sum_{r=0}^{\lfloor q/2
	\rfloor} \frac{1}{(2r)!} \int_{0}^{\infty}d\nu\; \nu^{2r+l-1}
	u^{(0)}(\sigma,\nu) \;\\
	&\qquad\times V_{l-1}(S^{l-1})\;
	(2r-1)!! \; C_{2r}\;
	 \delta_{a_{1}\dotsm a_{2r}}^{b_{1}\dotsm 
	b_{2r}}\; K^{a_{1}}{}_{b_{1}i_{1}}K^{a_{2}}{}_{b_{2}i_{1}} 
	\dotsm K^{a_{2r-1}}{}_{b_{2r-1}i_{r}} 
	K^{a_{2r}}{}_{b_{2r}i_{r}}\;.
\end{align*}
To perform the $S^{l-1}$ integral we used the averaging results derived in
Appendix~\ref{sec:avg-sphere}.  Next we use the Gauss equation
\begin{equation}
    R_{abcd}= K_{aci}K_{bdi}-K_{adi}K_{bci}
    \label{eq:Gauss-eq}
\end{equation}
to convert extrinsic curvature terms into intrinsic curvature terms:
\begin{align*}
    & \int_{\Sigma} \dual_{\Sigma} \sum_{r=0}^{\lfloor q/2 \rfloor}
    \frac{1}{2^{r}\, r!} \int_{0}^{\infty}d\nu\; \nu^{2r+l-1} u^{(0)}(\sigma,\nu)
    \;\\
    &\qquad\times V_{l-1}(S^{l-1})\; C_{2r}\; \frac{1}{2^{r}}\;
    \delta_{a_{1}\dotsm a_{2r}}^{b_{1}\dotsm b_{2r}}\;
    R_{a_{1}a_{2}b_{1}b_{2}} \dotsm
    R_{a_{2r-1}a_{2r}b_{2r-1}b_{2r}}\;.
\end{align*}
Rewrite the above using the curvature $2$-form $\Omega_{ab} = 
\frac{1}{2} R_{abcd} \theta^{c}\wedge \theta^{d}$ and identity 
\eqref{eq:basic-identity} to obtain
\begin{equation}
   E^{(0)} = V_{l-1}(S^{l-1}) \sum_{r=0}^{\lfloor q/2 \rfloor} C_{2r}
   \int_{\Sigma} \curv_{2r}(\Sigma)\; \dual_{\Sigma} \int_{0}^{\infty}
   d\nu\; \nu^{2r+l-1} \;u^{(0)}(\sigma,\nu)\,,
    \label{eq:foo}
\end{equation}
where
\begin{equation}
    \begin{split}
    \curv_{2r}(\Sigma)\; \dual_{\Sigma} &= \frac{1}{4^{r}\, r!}\; 
    \delta_{a_{1}\dotsm a_{2r}}^{b_{1}\dotsm
    b_{2r}}\; R_{a_{1}a_{2}b_{1}b_{2}} \dotsm
    R_{a_{2r-1}a_{2r}b_{2r-1}b_{2r}}\, \dual_{\Sigma}\,, \\
    &= \frac{1}{2^{r}\, r!}\; \dual^{a_{1}a_{2}\dotsm a_{2r-1}a_{2r}}
    \wedge \Omega_{a_{1}a_{2}} \wedge\dotsb \wedge
    \Omega_{a_{2r-1}a_{2r}}\,.
    \end{split}
    \label{eq:def-curv}
\end{equation}

Next we define the ``normal radial moments'' of the monopole moment of
the energy density by
\begin{equation}
    \mu^{(0)}_{2r}(\sigma) = \int_{(T_{\sigma}\Sigma)^{\perp}} 
    \lVert\bnu\rVert^{2r}\; 
    u^{(0)}(\sigma,\lVert\bnu\rVert)\; d^{l}\nu = V_{l-1}(S^{l-1}) \int_{0}^{\infty} 
    d\nu\; \nu^{2r+l-1}\; u^{(0)}(\sigma,\nu)\;.
    \label{eq:def-mu2r}
\end{equation}
Collating everything we  have the monopole contribution to the energy
\begin{equation}
    E^{(0)} = \sum_{r=0}^{\lfloor q/2 \rfloor} C_{2r}\; \int_{\Sigma} 
    \mu^{(0)}_{2r}(\sigma)\; \curv_{2r}(\Sigma)\; \dual_{\Sigma}\,.
    \label{eq:gen-Weyl}
\end{equation}

We now address the question of whether the finite series in
eq.~\eqref{eq:gen-Weyl} is exact or an approximation.  Assume the
integrable function $u^{(0)}$ is compactly supported in a nice
tube\footnote{The radius of the tube $\rho$ is small enough so that
there are are no self-intersections.} $\tube(\Sigma^{q},\rho)$.  In
this case, equation~\eqref{eq:gen-Weyl} is exact.  This is quite a
surprising result, and the main mathematical result of this paper
along with its generalization \eqref{eq:energy-total} to non
spherically symmetric functions.  It states that the integral of a
spherically symmetric compactly supported function may be described in
terms of a much smaller set of data: $(1 + \lfloor q/2 \rfloor)$
functions $\{ \mu^{(0)}_{2r}(\sigma) \}$ defined on $\Sigma$.  This
universal form tells us that only some general features of the energy
density function survive after integration.  Namely, a finite number
of radial moments.  The prime example of this integration result is
the Weyl volume formula~\eqref{eq:Weyl-tube-for-1}.  If the radius
$\rho$ is too large and there are self intersections.  then formula
\eqref{eq:gen-Weyl} will not be valid.  This can be seen by looking at
Figure~\ref{fig:tube-large-radius} and observing that certain volumes
will be over counted in attempting to perform the integration by first
integrating over the normal bundle.  This result for compactly
supported energyt densities motivates why there is a universality in
the type of effective field theories that are induced on $\Sigma^{q}$
from the ambient bulk physics.

In most physical applications, the energy density $u^{(0)}$ does not
have compact support.  In quantum field theories with massive
excitations you expect some type of exponential decay of the energy
density with a correlation length $\corr$ as you move away from
$\Sigma^{q}$:
\begin{equation}
    u^{(0)}(\sigma,\nu) \xrightarrow{\nu\to\infty\;} C 
    \nu^{-\alpha}\; e^{-\nu/\corr}\,,
\end{equation}
where $\alpha\ge 0$ in many models.  If $L \sim 1/\lVert \bK \rVert$
is characteristic of the distance at which you find the nearest self
intersection of the tube, and if you assume that $\corr < L$, then you
expect the there will be exponentially small corrections to
\eqref{eq:gen-Weyl} of order $e^{-L/\corr}$, note that $L/\corr \sim
1/(\corr\lVert \bK \rVert)$:
\begin{equation*}
    E^{(0)} = \sum_{r=0}^{\lfloor q/2 \rfloor} C_{2r}\; \int_{\Sigma} 
    \mu^{(0)}_{2r}(\sigma)\; \curv_{2r}(\Sigma)\; \dual_{\Sigma}
    + O\left( e^{-L/\corr}\right)\,.
\end{equation*}

Is there an  expansion parameter for the individual summands in
\eqref{eq:gen-Weyl}?  The answer to this question was essentially
given in the caption of Figure~\ref{fig:tube-large-radius}.  In the
context of our exponentially decaying energy density we would like for
$\corr \lVert \bK \rVert <1$.  We naively estimate the ratio of the
summands in the expansion.  First we assume that the couplings are
independent of $\sigma$ such as in the case of a spherically symmetric
defect, see Section~\ref{sec:approximate}.  If we think of $E^{(0)}$
as an effective dimensionless energy or effective action entering a
Boltzmann factor then the dimensions of $u^{(0)}$ are $[u^{(0)}] =
L^{-n} = M^{n}$.  For a static defect, the energy density is
translationally invariant along the defect and we expect from
dimensional analysis that $T_{q} \sim \mu_{n}^{n} \corr^{l}$, where
$\mu_{n}$ is some $n$-dimensional energy scale, and the
$q$-dimensional energy density $T_{q}$ will be identified with the
$p$-brane tension.  For example, $\corr$ could be a correlation length
in some quantum field theory that interacts with the submanifold
$\Sigma^{q}$.  Note the dimensional units of $T_{q}$ are
$[T_{q}]=L^{-q}= M^{q}$.  From eq.~\eqref{eq:def-mu2r} we see that
$\mu_{2r}^{(0)} \sim T_{q}\corr^{2r}$.  Since $\mu_{0}^{(0)}$ is the
coupling of the $q$-volume contribution, it is the $p$-brane tension
and we immediately have its identification with $T_{q}$.  Note that
the ratio of couplings $\mu_{2r+2}^{(0)}/\mu_{2r}^{(0)} \sim
\corr^{2}$.  The $q$-dimensional reciprocal Newtonian gravitational
constant $G_{q}^{-1} = (\MPl_{q})^{q-2} \sim \mu_{2}^{(0)} \sim
\corr^{2} T_{q}$.  There is a dependence on $l$ through the
coefficients $C_{2r}$ that we have not taken into account in our very
rough estimates.  The Gauss equation tells us that, roughly, $\lVert
\bR \rVert \sim \lVert \bK \rVert^{2}$.  An estimate of the $r$-th
summand in \eqref{eq:gen-Weyl} is given by
\begin{equation}
    \mu_{2r}^{(0)} \int_{\Sigma} \curv_{2r}(\Sigma)\; \dual_{\Sigma} 
    \sim \left( T_{q} \corr^{2r}\right) \lVert \bK \rVert^{2r} 
    \vol_{q}(\Sigma) \sim \left(\corr \lVert \bK \rVert \right)^{2r} 
    \left(T_{q} \vol_{q}(\Sigma)\right)\,.
    \label{eq:curv-estimate}
\end{equation}
Thus expansion parameter in this spherically symmetric energy density
model is $\left( \corr \lVert \bK \rVert\right)^{2} \sim \corr^{2}
\lVert \bR \rVert$.

\subsection{The dipole contribution}
\label{sec:dipole}

To evaluate the dipole moment contribution to the energy we begin with
equation \eqref{eq:W-int}, replace $u(\sigma,\bnu)$ by
$u^{(1)}_{i}(\sigma,\lVert \bnu \rVert)\, \nuhat^{i}$, and obtain
\begin{align*}
	& \int_{\Sigma} \dual_{\Sigma}\int_{(T_{\sigma}\Sigma)^{\perp}}
	u^{(1)}_{i}(\sigma,\lVert \bnu 
	\rVert)\, \nuhat^{i}\; \det(I +
	\bnu\bdot\bK)\; d^{l}\nu \\
	&= \int_{\Sigma} \dual_{\Sigma}\sum_{r=0}^{q}
	\int_{\nu=0}^{\infty}\int_{S^{l-1}} u^{(1)}_{i}(\sigma,\lVert
	\bnu \rVert) \; \frac{\nu^{r}}{r!}\;
	\delta_{a_{1}\dotsm a_{r}}^{b_{1}\dotsm b_{r}}\;
	K^{a_{1}}{}_{b_{1}i_{1}}K^{a_{2}}{}_{b_{2}i_{2}} \dotsm
	K^{a_{r}}{}_{b_{r}i_{r}}\\
	&\qquad\qquad \times \nuhat^{i}
	\nuhat^{i_{1}}\nuhat^{i_{2}}\dotsm \nuhat^{i_{r}} \cdot
	\nu^{l-1}\; d\nu\; d\vol_{S^{l-1}}\;,\\
	&= \int_{\Sigma} \dual_{\Sigma} \sum_{s=1}^{\lfloor (q+1)/2
	\rfloor} \frac{1}{(2s-1)!} \int_{0}^{\infty}d\nu\; 
	\nu^{2s+l-2}\;
	u^{(1)}_{i}(\sigma,\nu) \;\\
	&\qquad\times V_{l-1}(S^{l-1})\;
	C_{2s}\; \sym^{ii_{1}i_{2} \dotsm i_{2s-1}}\;
	 \delta_{a_{1}\dotsm a_{2s-1}}^{b_{1}\dotsm 
	b_{2s-1}}\; K^{a_{1}}{}_{b_{1}i_{1}}K^{a_{2}}{}_{b_{2}i_{2}} 
	\dotsm K^{a_{2s-1}}{}_{b_{2s-1}i_{2s-1}} \;,\\
	&= \int_{\Sigma} \dual_{\Sigma} \sum_{s=1}^{\lfloor (q+1)/2
	\rfloor} \frac{1}{(2s-1)!} \int_{0}^{\infty}d\nu\; 
	\nu^{2s+l-2}\;
	u^{(1)}_{i}(\sigma,\nu) \; V_{l-1}(S^{l-1})\; C_{2s}\; (2s-1)!! \\
	&\qquad\times 
	\delta_{a_{1}\dotsm a_{2s-1}}^{b_{1}\dotsm b_{2s-1}}\;
	K^{a_{1}}{}_{b_{1}i_{1}}K^{a_{2}}{}_{b_{2}i_{1}} \dotsm
	K^{a_{2s-3}}{}_{b_{2s-3}i_{s-1}}
	K^{a_{2s-2}}{}_{b_{2s-2}i_{s-1}}
	K^{a_{2s-1}}{}_{b_{2s-1}i}\;,\\
	&= \int_{\Sigma} \dual_{\Sigma} \sum_{s=1}^{\lfloor (q+1)/2
	\rfloor} \frac{1}{(2s-1)!} \int_{0}^{\infty}d\nu\; 
	\nu^{2s+l-2}\;
	u^{(1)}_{i}(\sigma,\nu) \; V_{l-1}(S^{l-1})\; C_{2s}\; (2s-1)!! \\
	&\qquad\times \frac{1}{2^{s-1}}\;
	\delta_{a_{1}\dotsm a_{2s-1}}^{b_{1}\dotsm b_{2s-1}}\;
	R_{a_{1}a_{2}b_{1}b_{2}} \dotsm
	R_{a_{2s-3}a_{2s-2}b_{2s-3}b_{2s-2}}
	K^{a_{2s-1}}{}_{b_{2s-1}i}\;,
\end{align*}
This may be expressed in terms of differential forms by introducing the
extrinsic curvature $1$-forms\footnote{The Riemannian connection
$\omega_{ai}$ in a Darboux frame adapted to $\Sigma$ is the same as
$\curvform_{ai}$.}
\begin{equation}
    \curvform_{ai} = K_{abi}\theta^{b}\,.
    \label{eq:def-curv-form}
\end{equation}
The complicated expression above becomes
\begin{align*}
    & \int_{\Sigma} \sum_{s=1}^{\lfloor (q+1)/2
    \rfloor} \frac{C_{2s}}{2^{s-1}\;(s-1)!} \int_{0}^{\infty}d\nu\; 
    \nu^{2s+l-2}\;
    u^{(1)}_{i}(\sigma,\nu) \; V_{l-1}(S^{l-1}) \\
    &\qquad\times
      \Omega_{a_{1}a_{2}}\wedge \dotsb \wedge 
      \Omega_{a_{2s-3}a_{2s-2}} \wedge \curvform_{a_{2s-1}i} \wedge 
      \dual^{a_{1}a_{2}\dotsm a_{2s-1}}\,.
\end{align*}
Mimicking \eqref{eq:def-mu2r} we define the normal radial
moments of the dipole moment of the energy density by
\begin{equation}
    \mu^{(1)}_{i, 2s-1}(\sigma) = \int_{(T_{\sigma}\Sigma)^{\perp}} 
    \lVert\bnu\rVert^{2s-1}\; 
    u^{(1)}_{i}(\sigma,\lVert\bnu\rVert)\; d^{l}\nu = V_{l-1}(S^{l-1}) \int_{0}^{\infty} 
    d\nu\; \nu^{2s+l-2}\; u^{(1)}_{i}(\sigma,\nu)\;.
    \label{eq:def-mu1}
\end{equation}
Putting it all together we see that the dipole contribution to the 
energy is given by
\begin{equation}
    E^{(1)} = \sum_{s=0}^{\lfloor (q-1)/2 \rfloor}
    \frac{C_{2s+2}}{2^{s}\;s!} 
    \int_{\Sigma} \mu^{(1)}_{i}{}_{2s+1}(\sigma)\;
    \curvform_{bi} \wedge \Omega_{a_{1}a_{2}}\wedge \dotsb \wedge
    \Omega_{a_{2s-1}a_{2s}} \wedge \dual^{ba_{1}a_{2}\dotsm
    a_{2s}}\,.
    \label{eq:energy-dipole-1}
\end{equation}

\subsection{The general multipole contribution}%
\label{sec:general-multipole}

The contribution to the energy from the cartesian $2^{j}$-pole
is given by (see \eqref{eq:W-int})
\begin{equation}
    \label{eq:multienergy}
\begin{split}
    & \int_{\Sigma} \dual_{\Sigma}\int_{(T_{\sigma}\Sigma)^{\perp}}
    u^{(j)}_{k_{1}\dotsm k_{j}}(\sigma, \lVert\bnu\rVert)\; 
    \nuhat^{k_{1}}\nuhat^{k_{2}}\dotsm \nuhat^{k_{j}}
    \; \det(I +
    \bnu\bdot\bK)\; d^{l}\nu \\
    &= \int_{\Sigma} \dual_{\Sigma}\sum_{r=0}^{q}
    \int_{\nu=0}^{\infty}\int_{S^{l-1}} 
    u^{(j)}_{k_{1}\dotsm k_{j}}(\sigma, \nu)\;
    \frac{\nu^{r}}{r!}\; \delta_{a_{1}\dotsm a_{r}}^{b_{1}\dotsm
    b_{r}}\; K^{a_{1}}{}_{b_{1}i_{1}}(\sigma)
    K^{a_{2}}{}_{b_{2}i_{2}}(\sigma) \dotsm
    K^{a_{r}}{}_{b_{r}i_{r}}(\sigma)\\
    &\qquad\qquad 
    \times \nuhat^{k_{1}}\nuhat^{k_{2}}\dotsm \nuhat^{k_{j}}
    \times
    \nuhat^{i_{1}}\nuhat^{i_{2}}\dotsm \nuhat^{i_{r}} \cdot
    \nu^{l-1}\; d\nu\; d\vol_{S^{l-1}}\;.
\end{split}
\end{equation}
The spherical integral vanishes unless $j+r$ is an even number.
The expression above may be rewritten as
\begin{align*}
    & \int_{\Sigma} \dual_{\Sigma}\sum_{r\in \mathcal{R}}
    \int_{0}^{\infty}d\nu\; 
    u^{(j)}_{k_{1}\dotsm k_{j}}(\sigma, \nu)\;
    \frac{\nu^{r+l-1}}{r!}\; \delta_{a_{1}\dotsm a_{r}}^{b_{1}\dotsm
    b_{r}}\; K^{a_{1}}{}_{b_{1}i_{1}}(\sigma)
    K^{a_{2}}{}_{b_{2}i_{2}}(\sigma) \dotsm
    K^{a_{r}}{}_{b_{r}i_{r}}(\sigma)\\
    &\qquad\qquad \times V_{l-1}(S^{l-1})\; C_{j+r}\; \sym^{k_{1}\dotsm
    k_{j}i_{1}\dotsm i_{r}}\,,
\end{align*}
where the summation set $\mathcal{R}$ will be specified shortly.
$\sym^{k_{1}\dotsm k_{j}i_{1}\dotsm i_{r}}$ is a sum of $(j+r-1)!!$
summands constructed from Kronecker $\delta$-symbols, see
eq.~\eqref{eq:wick}.  A summand that contains $\delta^{kk'}$
contributes zero to the sum because $u^{(j)}_{k_{1}\dotsm k_{j}}$ is
traceless.  Hence each ``$k$'' index must be contracted with an
``$i$'' index to obtain a non-zero contribution.  Thus we conclude
that non-vanishing terms must have $r - j = 2s$ where $ 0 \le s \le
\lfloor (q-j)/2 \rfloor$.  Inserting this information into the
displayed equation above yields
\begin{equation}
    \begin{split}
    & \int_{\Sigma} \dual_{\Sigma}\sum_{s=0}^{\lfloor(q-j)/2\rfloor}
    \int_{0}^{\infty}d\nu\; u^{(j)}_{k_{1}\dotsm k_{j}}(\sigma,
    \nu)\; \frac{\nu^{2s+j+l-1}}{(j+2s)!}\; V_{l-1}(S^{l-1})\;
    C_{2(j+s)} \\
    &\qquad\qquad \times \; \sym^{k_{1}\dotsm
    k_{j}i_{1}\dotsm i_{j+2s}} \; \delta_{a_{1}\dotsm a_{j+2s}}^{b_{1}\dotsm
    b_{j+2s}}\; K^{a_{1}}{}_{b_{1}i_{1}}(\sigma)
    K^{a_{2}}{}_{b_{2}i_{2}}(\sigma) \dotsm
    K^{a_{r}}{}_{b_{j+2s}i_{j+2s}}(\sigma)\,.
    \end{split}
    \label{eq:energy-j-1}
\end{equation}
Remember that the total number of indices in $\sym$ is $j+r = 2(j+s)$.
The number of summands in $\sym^{k_{1}\dotsm k_{j}i_{1}\dotsm i_{r}}$
that give a non-vanishing contribution is
\begin{equation}
    \begin{split}
    \underbrace{(j+2s)(j+2s-1) \dotsm (2s+1)}_{\text{number
    of contractions of type ``$ik$''}} \times (2s-1)!!  &=
    \frac{(j+2s)!}{(2s)!}\times (2s-1)!!\,,\\
    &= \frac{(j+2s)!}{2^{s}\; s!}\;.
    \end{split}
    \label{eq:combinatoric}
\end{equation}
Define the radial moments of the $2^{j}$-pole by
\begin{equation}
    \begin{split}
    \mu^{(j)}_{k_{1}\dotsm k_{j}}{}_{,2s+j}(\sigma) &=
    \int_{(T_{\sigma}\Sigma)^{\perp}} \lVert \bnu \rVert^{2s+j} \;
    u^{(j)}_{k_{1}\dotsm k_{j}}(\sigma, \lVert \bnu \rVert)\; d^{l}\nu
    \,, \\
    &= V_{l-1}(S^{l-1})\int_{0}^{\infty}d\nu\; \nu^{2s+j+l-1}\;
    u^{(j)}_{k_{1}\dotsm k_{j}}(\sigma, \nu)\,.
    \end{split}
    \label{eq:radial-mom-j}
\end{equation}
Note that the moments $\mu^{(j)}_{k_{1}\dotsm
k_{j}}{}_{,2s+j}(\sigma)$ are a section of the vector bundle (over
$\Sigma$) which is the symmetric traceless subspace of the $j$-th
tensor product of $(T\Sigma)^{\perp}$.

Inserting the radial moments definitions \eqref{eq:radial-mom-j} into
\eqref{eq:multienergy} and using the Gauss equation to rewrite some of
the factors as the intrinsic curvature leads to
\begin{align*}
    \sum_{s=0}^{\lfloor(q-j)/2\rfloor} \frac{C_{2j+2s}}{2^{s}\; s!}
    & \int_{\Sigma} \dual_{\Sigma}\; \mu^{(j)}_{k_{1}\dotsm
    k_{j}}{}_{,2s+j}(\sigma) \; \delta_{a_{1}\dotsm
    a_{j+2s}}^{b_{1}\dotsm b_{j+2s}}\\
    &\quad\times K^{a_{1}}{}_{b_{1}k_{1}}(\sigma)
    K^{a_{2}}{}_{b_{2}k_{2}}(\sigma) \dotsm
    K^{a_{j}}{}_{b_{j}k_{j}}(\sigma) \\
    &\quad \times 
    K^{a_{j+1}}{}_{b_{j+1}i_{1}}(\sigma)
    K^{a_{j+2}}{}_{b_{j+2}i_{1}}(\sigma) \dotsm
    K^{a_{j+2s-1}}{}_{b_{j+2s-1}i_{s}}(\sigma)
    K^{a_{j+2s}}{}_{b_{j+2s}i_{s}}(\sigma) \,,\\
    &= \sum_{s=0}^{\lfloor(q-j)/2\rfloor} \frac{C_{2j+2s}}{2^{s}\; s!}
     \int_{\Sigma} \dual_{\Sigma}\; \mu^{(j)}_{k_{1}\dotsm
    k_{j}}{}_{,2s+j}(\sigma) \; \delta_{a_{1}\dotsm
    a_{j+2s}}^{b_{1}\dotsm b_{j+2s}}\\
    &\quad\times K^{a_{1}}{}_{b_{1}k_{1}}(\sigma)
    K^{a_{2}}{}_{b_{2}k_{2}}(\sigma) \dotsm
    K^{a_{j}}{}_{b_{j}k_{j}}(\sigma) \\
    &\quad \times \frac{1}{2^{s}}\;
    R^{a_{j+1}a_{j+2}}{}_{b_{j+1}b_{j+2}} \dots
    R^{a_{j+2s-1}a_{j+2s}}{}_{b_{j+2s-1}b_{j+2s}}\;.
\end{align*}
The last expression above may be rewritten using differential forms 
and we obtain the following expression for the contribution of the 
$2^{j}$-pole to the energy
\begin{multline}
	E^{(j)}  = \sum_{s=0}^{\lfloor(q-j)/2\rfloor} 
	\frac{C_{2j+2s}}{2^{s}\; s!}   	\\
	\times \int_{\Sigma} \mu^{(j)}_{k_{1}\dotsm
	k_{j}}{}_{,2s+j}(\sigma) \; 
	\kappa_{b_{1}}{}^{k_{1}} \wedge
	\dotsb \wedge \kappa_{b_{j}}{}^{k_{j}}
	\wedge
	\Omega_{a_{1}a_{2}} \wedge \dotsb \wedge
	\Omega_{a_{2s-1}a_{2s}} \wedge \dual^{b_{1}\dotsm b_{j} a_{1}
	\dotsm a_{2s}}\,.
    \label{eq:energy-j-moment}
\end{multline}
As a check you can verify that the expression above reduces to
\eqref{eq:gen-Weyl} in the monopole case and
\eqref{eq:energy-dipole-1} in the dipole case.  Notice that if $q-j$
is even then the moments that occur are $\mu^{(j)}_{\bullet,j},
\mu^{(j)}_{\bullet,j+1}, \dotsc, \mu^{(j)}_{\bullet,q}$; and
if $q-j$ is odd you get $\mu^{(j)}_{\bullet,j},
\mu^{(j)}_{\bullet,j+1}, \dotsc, \mu^{(j)}_{\bullet,q-1}$.

The Gauss equation \eqref{eq:Gauss-eq} may be written as $\Omega_{ab}
= \curvform_{a}{}^{k} \wedge \curvform_{b}{}^{k}$ and the
$\SO(l)$-curvature $2$-form of the normal bundle is $F^{ij} =
\curvform_{a}{}^{i}\wedge \curvform_{a}{}^{j}$ (the dual Gauss
equation).  Since the cartesian multipole moments
$\mu^{(j)}_{k_{1}\dotsm k_{j}}$ are traceless in the $k$ indices, we
see that the $\kappa$ terms in \eqref{eq:energy-j-moment} cannot be
transformed into terms involving the intrinsic curvature $R_{abcd}$ of
the surface.  Note that the curvature $F^{ij}$ of the normal bundle
does not appear.  For completeness, we note the Codazzi-Mainardi
equation $ 0 = D\kappa_{ai} = d\kappa_{ai} +
\omega_{ab}\wedge\kappa_{bi} + \omega_{ij}\wedge\kappa_{aj}$.

Now we can write down a multipole expansion formula for the total
energy of an energy tube:
\begin{multline}
    E  = \sum_{j=0}^{q} \sum_{s=0}^{\lfloor(q-j)/2\rfloor} 
    \frac{C_{2j+2s}}{2^{s}\; s!}   	\\
    \times  \int_{\Sigma} 
    \mu^{(j)}_{k_{1}\dotsm k_{j}}{}_{,2s+j}(\sigma) \;
    \left(\kappa_{b_{1}}{}^{k_{1}} \wedge \dotsb \wedge
    \kappa_{b_{j}}{}^{k_{j}}\right)^{\text{ST}} \wedge
    \Omega_{a_{1}a_{2}} \wedge \dotsb \wedge \Omega_{a_{2s-1}a_{2s}}
    \wedge \dual^{b_{1}\dotsm b_{j} a_{1} \dotsm a_{2s}} 
    \label{eq:energy-total}
\end{multline}
In the above, the superscript $\text{ST}$ means orthogonal projection
onto the symmetric traceless part on the $k$-indices.
The total number of moments is $\sum_{j=0}^{q}\left(
\lfloor(q-j)/2\rfloor +1\right) \dim W^{j}$ where $\dim W^{j}$ is a
function of $l=n-q$.  It is quite interesting that integral
\eqref{eq:total-energy} is given by the finite number of terms in 
\eqref{eq:energy-total}.

We now repeat an earlier discussion in the spherically symmetric
case.  Is the the finite series in eq.~\eqref{eq:energy-total} exact
or an approximation?  Assume the integrable function $u$ is compactly
supported in a nice tube\footnote{The radius of the tube $\rho$ is
small enough so that there are are no self-intersections.}
$\tube(\Sigma^{q},\rho)$.  In this case,
equation~\eqref{eq:energy-total} is exactly the value of the energy
density integral.  If the radius $\rho$ is too large and there are
self intersections then formula \eqref{eq:energy-total} will not be
valid.  If then energy density decays exponentially then we expect
exponentially small corrections to \eqref{eq:energy-total} as in the
spherically symmetric case.

If $\mu^{(j)}_{\bullet,2s+j}$ gets a vacuum expectation value
(VEV), the structure group $\SO(l)$ of the normal bundle is reduced to
a subgroup that leaves the VEV invariant.  In this way you could have
some type of Nambu-Goldstone or Higgs mechanism on $\Sigma$. 

The expansion parameter for \eqref{eq:energy-total} is $\corr \lVert
\bK \rVert$ due to the presence of the extrinsic curvature terms
rather than $(\corr \lVert \bK \rVert)^{2}$ in the spherically
symmetric case \eqref{eq:gen-Weyl}.

\section{Embeddings and emergent theories of gravity}
\label{sec:embedding}

We begin with a differentiable $q$-manifold $\Stil = \Stil^{q}$ and
we are interested in embedding\footnote{We ignore the technical
differences between an embedding and an immersion.} $\Stil$ in
Euclidean $n$-space $\bbE^{n}$.  The reason for embedding is that we
will assume that there is a quantum field theory (QFT) on $\bbE^{n}$
and we are interested in discovering the effect of the interaction of
the QFT with the embedded submanifold. The role of the QFT is to 
provide a localized energy density near $\Sigma^{q}$. In this section 
we discuss the dynamical equations that arise when we vary the 
embedding. In Section~\ref{sec:defects} we address additional 
dynamical equations that arise due to variations in the energy 
density near $\Sigma^{q}$.

Before exploring the consequences of the embedding we digress and
explain what will not be considered in this paper.  The manifold
$\Stil$ could be endowed with intrinsic geometrical structures that
are not related to the embedding.  For example, assume $\Stil$ has a
Riemannian metric $\bgtil$.  We expect the action that determines the
dynamics of $\Stil$ to be of the form $\int_{\Stil} \left(a + b\,
R(\bgtil)+ \dotsb \right)\sqrt{\tilde{g}}\;d^{q}\sigma$ if $\bgtil$ is
a dynamical field.  In such a situation we see that we will have
$q$-dimensional gravitation on $\Stil$ \emph{a priori} of the
embedding.  We will not discuss this case at all, see for example
\cite{Dvali:2000hr}.  The problem we address is the one where $\Stil$
is a plain differentiable manifold with no intrinsic structures and
its geometry is induced by an embedding $\Stil$ in an Euclidean space
with a bulk QFT. We address the question whether potentially an effective
theory of $q$-dimensional gravity on $\Stil$ emerges because of the
embedding.  There is no fundamental graviton in the Euclidean space
$\bbE^{n}$ but the dynamics of the submanifold can effectively be
described by a gravity-like theory that only lives on the 
submanifold. In this section we discuss the meaning of \emph{gravity-like}.

It is worthwhile to  be mathematically precise to better understand
the goals of this section.  An embedding is given by a map $X:
\Stil \to \bbE^{n}$.  We denote the embedded submanifold by $\Sigma =
X(\Stil)$.  Note that $\Sigma \subset \bbE^{n}$ and thus we have an
inclusion map $\iota: \Sigma \hookrightarrow \bbE^{n}$.  If
$\eucl$ is the Euclidean metric on $\bbE^{n}$ then the induced
metric on $\Sigma$ given by the pullback
$\bm{g}=\iota^{*}\eucl$.  This induced metric on $\Sigma$ may be
viewed as a metric $\bgtil$ on $\Stil$ by pulling back again: $\bgtil
= X^{*}(\bm{g})=X^{*}(\iota^{*}\eucl) = (\iota \circ X)^{*}
\eucl$, see equation~\eqref{eq:def-g}.

If $(\Stil,\bgtil)$ is $q$-manifold with an intrinsic metric $\bgtil$
then $X: \Stil \to \bbE^{n}$ is an \emph{isometric embedding} if
$X^{*}\bm{g}= \bgtil$.  To truly have a theory of gravity, it is
necessary that you obtain all admissible intrinsic metrics on $\Sigma$
via embeddings.  For this to occur, various general theorems impose
restrictions on $n$, the dimensionality of the embedding space.

The Nash embedding theorem and its refinements, see
\cite{Greene:isometric,Clarke:embedding,Berger:PNAS,Bryant-Griffiths-Yang:isometric}
and references therein, provide bounds that state that a smooth
($C^{\infty}$) isometric embedding of a riemannian manifold
$\Sigma^{q}$ in euclidean space $\bbE^{n}$ exists locally if $n \ge
\frac{1}{2} q(q+1) + q$ and globally if $n \ge \frac{1}{2}q(q+1) +
3q+5$.  For real analytic ($C^{\omega}$) data, the isometric embedding
theorem of Burstin-Cartan-Janet-Schl\"{a}fly
\cite{Berger:PNAS,Berger:big} states that a local real analytic
embedding exists if $n \ge \frac{1}{2} q(q+1)$.  It is believed that
the proven local $C^{\infty}$ bounds are not optimal and that the
local $C^{\infty}$ threshold\footnote{See the discussion in Terry
Tao's notes about P.~Griffiths' work and comments by D.~Yang in
\href{http://terrytao.wordpress.com/2014/08/13/khot-osher-griffiths/}
{http://terrytao.wordpress.com/2014/08/13/khot-osher-griffiths}.
S.T.~Yau informs me that a motivation for the $C^{\infty}$ conjecture
is that mathematicians have looked very hard but have not found a
counterexample.} is actually $\half q(q+1)$.

The theorems discussed in the previous paragraph apply to a generic
manifold.  For very special manifolds, the bounds are smaller.  For
example a $q$ dimensional vector space with the Euclidean metric can
be globally isometrically embedded in any $\bbE^{n}$ with $n \ge q$.
The unit sphere $S^{q}$ with the round metric can be globally
isometrically embedded in any $\bbE^{n}$ with $n \ge q+1$.

\begin{remark}
    The dimension bound for a local analytic isometric embedding of a
    Lorentzian manifold $\Sigma^{q}$ in Minkowski space $\bbM^{n}$ is
    also $n \ge \half q(q+1)$ according to
    Eisenhart~\cite[p.~188]{Eisenhart:RG}.  If a $q$-manifold $\Sigma$
    has metric with signature $(q_{-},q_{+})$ where $q=q_{-}+ q_{+}$,
    and if you would like to locally analytically embed isometrically
    in $\mathbb{R}^n$ with signature $(n_{-},n_{+})$ where $n = n_- +
    n_+$, then you need $n \ge \frac{1}{2} q(q+1)$, and $n_{-}$ and
    $n_{+}$ constrained by $n_+ \ge q_+$, and $n_- \ge q_-$.  Global
    embedding theorems for Lorentzian manifolds analogous to the Nash
    theorems are discussed in
    Greene~\cite{Greene:isometric}, and in Clarke~\cite{Clarke:embedding}.
\end{remark}

The induced metric\footnote{From now on we think of the embedding as
being an isometric embedding and we identify the induced metric
$\bm{g}$ from the embedding with an intrinsic metric $\bgtil$.} is
given by $\bm{g} = d\bX \bdot d\bX$.  If we vary the map $X: \Stil \to
\bbE^{n}$ then the change in the induced metric is 
\begin{equation}
    \delta \bm{g} = d\bX \bdot d(\delta\bX) + d(\delta \bX)\bdot 
    d\bX\,,
    \label{eq:delta-g}
\end{equation}
where $\delta \bX$ is the variation of the map, a $1$-form on the
space of maps $\Map(\Stil,\bbE^{n})$.  Next we express $\delta \bX$
in terms of a deformation tangential to the surface and a
deformation orthogonal to the surface\footnote{See
Section~\ref{sec:vol-element} for the notation.} using an adapted 
Darboux frame
\begin{equation}
    \delta\bX =\ehat_{a} \xi^{a} + \nhat_{i} \psi^{i}\,,
    \label{eq:def-delta-X}
\end{equation}
where $\xi$ and $\psi$ are $1$-forms\footnote{The condition
$\delta^{2}\bX=0$ imposes some constrains on the differentials of the
$1$-forms $\xi$ and $\psi$: $\delta\xi + \omega_{\parallel}\wedge \xi
+ \kappa\wedge\psi=0$ and $\delta\psi + \omega_{\perp}\wedge\psi -
\kappa\wedge\xi=0$.} on the space of maps $\Map(\Stil,X)$.  Inserting
this decomposition into \eqref{eq:delta-g} we see that
\begin{align}
    \delta \bm{g}& = \theta^{a} \otimes D\xi^{a} + D\xi^{a} \otimes
    \theta^{a} + 2 K_{ab}{}^{i} \psi^{i}\, \theta^{a} \otimes
    \theta^{b}\,,
    \nonumber
    \\
    & = \left( D_{a}\xi_{b} + D_{b}\xi_{a} + 2 K_{ab}{}^{i}\psi^{i} 
    \right) \theta^{a} \otimes \theta^{b}\,,
    \label{eq:delta-g-2}
\end{align}
where $D\xi^{a} = d\xi^{a} + \omega_{ab}\xi^{b}$.  As expected, the
tangential projection $\bm{\xi}$ of the deformation is a vector field
along the surface and thus the variation of the metric contains a part
that is an infinitesimal diffeomorphism given by the Lie derivative
$\Lie_{\xi}g_{ab} = -(D_{a}\xi_{b} + D_{b}\xi_{a})$.
Equation~\eqref{eq:delta-g-2} is the differential of the map given by
\eqref{eq:def-g}.  What are the conditions that the differential map
\eqref{eq:delta-g-2} be surjective?  On the left hand side of the
equation, we have $\frac{1}{2} q(q+1)$ functions $\delta g_{ab}$ on
$\Sigma$; the Lie derivative term on the right hand side involves $q$
functions $\xi^{a}$ on $\Sigma$.  For surjectiveness of the
differential map, we require that the map $\mathcal{K}:
T_{\sigma}\Sigma^{\perp} \to \Sym^{2}(T_{\sigma}\Sigma)$ given by
$\mathcal{K}: \psi \mapsto K_{ab}{}^{i}\psi^{i}$ have at least linear
transformation rank\footnote{The word \emph{rank} is used in two
different senses in this paper: firstly, in the linear algebraic sense
of the rank of a linear transformation; secondly, in the sense of the
rank of a vector bundle, \emph{i.e.}, the dimensionality of the
fiber.} $\frac{1}{2} q(q+1) - q = \frac{1}{2} q(q-1)$.  This means
that, in the vector bundle sense, $\rank (T\Sigma^{\perp}) \ge
\frac{1}{2}q(q-1)$.  For surjectiveness of the differential map, we
need that the dimension of the embedding space satisfy $ n=q+l \ge
\frac{1}{2} q(q+1)$.  This agrees with the naive counting in PDE
system \eqref{eq:def-g} consisting of $\half q(q+1)$ first order PDE
for the $n$ embedding functions $X^{\mu}$.  To obtain an isometric
embedding you need at least $\half q(q+1)$ functions $X^{\mu}$ to
naively avoid having an over determined system of PDE.

If you take a general Lovelock action~\eqref{eq:gen-Weyl-x} and you
vary the embedding you will get
\begin{subequations}\label{eq:var-love}
\begin{align}
    \delta I &= -\half \sum_{r=0}^{\lfloor q/2 \rfloor} C_{2r}
    \mu^{(0)}_{2r} \int_{\Sigma} \sqrt{\det g} \, \love{2r}^{ab}\; (\delta
    g_{ab})\; d^{q}\sigma
    \label{eq:var-love-1}\\
    & = -\half \sum_{r=0}^{\lfloor q/2 \rfloor} C_{2r}
    \mu^{(0)}_{2r} \int_{\Sigma} \sqrt{\det g} \, \love{2r}^{ab}
    \left( D_{a}\xi_{b} + D_{b}\xi_{a} + 2 K_{ab}{}^{i}\psi^{i} 
    \right)\; d^{q}\sigma \,.
    \label{eq:var-love-2} 
\end{align}
\end{subequations}
In the above, $\love{2r}^{ab}$ are the conserved Lovelock
tensors~\eqref{eq:def-love-2r}.  The equation of motion that you get 
from the tangential variations is
\begin{equation}
    0 = \sum_{r=0}^{\lfloor q/2 \rfloor} C_{2r} 
    \mu^{(0)}_{2r}\, D_{a}\love{2r}^{ab}\,.
    \label{eq:eom-tangential}
\end{equation}
This equation is a reflection of the $\Diff_{0}(\Sigma)$ invariance of
the action.  In fact, since each summand in \eqref{eq:gen-Weyl-x} is
$\Diff_{0}(\Sigma)$ invariant, each $D_{a}\love{2r}^{ab}=0$
identically.  The equations of motion arising from varying the
embedding in directions normal to the surface are
\begin{equation}
    0 = \left(\sum_{r=0}^{\lfloor q/2 \rfloor} C_{2r} 
    \mu^{(0)}_{2r}\, \love{2r}^{ab} \right) K_{ab}{}^{i}
    \label{eq:eom-normal}
\end{equation}
There are two cases to consider. 

First, assume that $n \ge \half 
q(q+1)$. Our previous discussion shows that we get all possible 
metric variations $\delta g_{ab}$, and we can immediately use 
\eqref{eq:var-love-1}, and conclude that the equations of motions are
\begin{equation}
    0 = \sum_{r=0}^{\lfloor q/2 \rfloor} C_{2r} 
	\mu^{(0)}_{2r}\, \love{2r}^{ab}\,.
    \label{eq:eom-large}
\end{equation}
The equations of motion for the dynamics of $\Sigma^{q}$ are those of
an euclidean Lovelock theory of gravity.   Classically, this
looks like an emergent theory of gravity where there is a
graviton-like excitation on the surface.  There are no negative metric
graviton states.  Note that these equations just involve intrinsic 
geometrical data on $\Sigma^{q}$ and are not aware of the embedding. 
We remind the reader that there may be additional equations that 
arise from the fields in the surrounding QFT and these are discussed 
in Section~\ref{sec:defects}.

The next case is where $n < \half q(q+1)$.  In this case you do not
expect that all possible deformations of the surface lead to all
allowed intrinsic metrics on the surface.  Here you cannot use
\eqref{eq:var-love-1} directly but must use \eqref{eq:var-love-2} that
tells you which variations of the metric you obtain by varying the
embedding.  The equations of motions that follow are the tautological
equations \eqref{eq:eom-tangential} and a subset of the Lovelock
equations given by \eqref{eq:eom-normal}.  You get a gravity-like
theory but it is not gravity in the sense that the excitations are not
gravitons as we explain below.

We study \eqref{eq:eom-normal} the weak field linearized
approximation.  By the remarks in
Zumino~\cite{Zumino:1985dp}, the only linearized term that is
dynamical is contained in the ordinary Einstein tensor
$\love{2}^{ab}$.  If we write the weak field metric on the 
surface\footnote{For the remaining part of this section we denote the 
metric on $\Sigma^{q}$ as $\bm{h} = h_{ab}\; d\sigma^{a} \otimes 
d\sigma^{b}$.} as
$h_{ab} = \delta_{ab} + \hpert_{ab}$, and if we define the auxiliary
variables $\hpertb_{ab} = \hpert_{ab} - \half
\delta_{ab}\hpert^{c}{}_{c}$, then
\begin{equation}
    \love{2}^{ab}= \half \left(-\partial^{2}\, \hpertb^{ab} +
    \partial^{a}\partial_{c}\hpertb^{bc} +
    \partial^{b}\partial_{c}\hpertb^{ac} - 
    \delta^{ab}\partial_{c}\partial_{d} \hpertb^{cd}
    \right)\,.
    \label{eq:einst-lin}
\end{equation}
The linearized Einstein tensor is gauge invariant under the linearized
gauge transformation $\hpert_{ab} \to \hpert_{ab} +
\partial_{a}\xi_{b} + \partial_{b}\xi_{a}$, or $\hpertb_{ab} \to
\hpertb_{ab} + \partial_{a}\xi_{b} +
\partial_{b}\xi_{a}-\delta_{ab}\partial_{c}\xi^{c}$.  Next we use a
gauge transformation and gauge fix in Lorenz gauge
$\partial_{c}\hpertb^{cd}=0$.  This imposes $q$ conditions on the
$\half q(q+1)$ functions $\hpertb_{ab}$ and leaves us with $\half
q(q-1)$ functions that encapsulate the Euclidean degrees of freedom.
Imposing Lorenz gauge on \eqref{eq:einst-lin} leads to
\begin{equation}
    \love{2}^{ab}= -\half \partial^{2} \hpertb^{ab},
    \label{eq:einst-lin-gf}
\end{equation}
where there are only $\half q(q-1)$ independent functions
$\hpertb_{ab}$.  These manipulations only require properties of the
linearized Einstein tensor.

If $l \ge \half q(q-1)$, then we are in situation
\eqref{eq:eom-large}, and the dynamics of $\Sigma^{q}$ are the
dynamics of gravity governed by the wave equations
\eqref{eq:einst-lin-gf}.

If $l < \half q(q-1)$, the dynamics of $\Sigma^{q}$ are 
described by  \eqref{eq:eom-normal}. As far as counting degrees of 
freedom, we can treat $K_{ab}{}^{i}$ as constants and the linearized 
equation of motion in Lorenz gauge becomes
\begin{equation}
    \partial^{2}\, \chi^{i} = 0 \quad\text{where } \chi^{i} = 
    \hpertb^{ab}\, K_{ab}{}^{i}\,.
    \label{eq:def-chi}
\end{equation}
Generically you expect the map\footnote{The map $\mathcal{K}^{*}$ is
essentially the adjoint of the map $\mathcal{K}$ previously
discussed.} $\mathcal{K}^{*}:\hpertb \mapsto \chi$ to be surjective in
the case with $\half q(q-1) > l$.  This means that there is a
non-trivial kernel with $\dim\ker \mathcal{K}^{*}=\half q(q-1) - l \ge
1$.  Thus we have $l$ dynamical fields $\phi^{i}$ with $l <\half
q(q-1)$ that satisfy \eqref{eq:def-chi}, and there are
$\dim\mathcal{K}^{*}$ linear combinations of the metric fluctuations
$\hpertb$ that vanish automatically and do not satisfy the Laplace
equation.  The metric perturbations in the kernel of $\mathcal{K}^{*}$
are not dynamical.  The number of degrees of freedom of this theory are
less than the number of degrees of freedom in a gravitational theory.
On the other hand, the degrees of freedom here, $\chi^{i} =
\hpertb^{ab}\, K_{ab}{}^{i}$, are linear combinations of the metric
fluctuations, and in this sense the theory is \emph{gravity-like}.
The properties of these gravity-like theories should be explored.

\begin{remark}
    Note that $\sqrt{\det h}\; h^{ab} \approx \delta^{ab}
    -\hpertb^{ab}$, and therefore we conjecture that the dynamical
    degrees of freedom beyond the weak field approximation in the case
    $l < \half q(q-1)$ are the mean curvature vector density,
    $\sqrt{\det h}\; h^{ab}\; K_{ab}{}^{i}$.
\end{remark}
\begin{remark}
    The counting of degrees of freedom is different between gauge
    theories in Minkowski space and in Euclidean space.  In Minkowski
    space folklore ``each diffeomorphism gauge transformation kills
    twice''.  The reason is that once the Lorenz gauge condition
    $\partial_{c}\hpertb^{cd}=0$ is imposed, it can be maintained by
    performing an additional gauge transformation that satisfies
    $\partial^{2}\xi_{a}=0$.  In Euclidean space there are no
    acceptable harmonic functions $\xi_{a}$.  But in Minkowski space,
    you are solving the wave equation because $\partial^{2}$ is the
    dalembertian wave operator, and there are acceptable solutions
    $\xi_{a}$ to the wave equation.  This allows you to impose an
    additional $q$ conditions on $\hpertb_{ab}$ and get down to $\half
    q(q+1) -2 \cdot q = \half q(q-3)$ Minkowski physical degrees of
    freedom.  This is the dimensionality of the irreducible symmetric
    traceless representation of $\SO(q-2)$ which is the compact
    subgroup of the Wigner little group for a massless particle in
    $\bbM^{q}$.
\end{remark}

Finally, we briefly remark on the difference between the ``kinematics''
of embeddings and imposing a ``dynamics'' on an embedding.  If
$\sigma^{a}$ are coordinates on $\Sigma$ and if we work in a
coordinate frame for $T\Sigma$, then the coordinate basis vectors
$\evec_{a}= \partial/\partial\sigma^{a}$ are given by
\begin{equation}
    \partial_{a}\bX = \bE_{\mu}\; \partial_{a}X^{\mu}= \evec_{a} \,.
    \label{eq:def-evec}
\end{equation}
By taking the exterior derivative  we find
\begin{equation*}
    \bE_{\mu}\; \partial_{b}(\partial_{a} X^{\mu})\; d\sigma^{b} =
    d\evec_{a} = \evec_{b}\; \Gamma_{c}{}^{b}{}_{a}\; d\sigma^{c}
    - \nhat_{j}\; K_{ab}{}^{j}\; d\sigma^{b}\,,
\end{equation*}
where we applied the coordinate basis version of \eqref{eq:def-de}.
Rewriting the above we obtain
\begin{equation}
    D_{b}(\partial_{a}\bX) = 
    \bE_{\mu}\; D_{b}(\partial_{a}X^{\mu}) = - \nhat_{j} \;
    K_{ab}{}^{j}\,,
    \label{eq:hessian-X}
\end{equation}
where $D_{a}$ is the covariant derivative on $T\Sigma$.  Note that
equation \eqref{eq:hessian-X} is a \emph{kinematic result}; it is
equivalent to the definition of the extrinsic curvature we gave in
\eqref{eq:def-dbasis}.  Typically, the dynamics are determined by
imposing constraints on the second partial derivatives.  You expect
the dynamics to involve the laplacian\footnote{More properly, the
dalembertian in a Lorentzian framework.} of $X$ and we see that
$D^{a}(\partial_{a} \bX) = - \nhat_{j}\; K^{a}{}_{a}{}^{j}$.  The
laplacian of the embedding map $\bX$ is the mean curvature vector
$\nhat_{j}\; K^{a}{}_{a}{}^{j}$.  The map $X:\Stil \to \bbE^{n}$ is
harmonic if and only if the submanifold $X(\Stil) =\Sigma
\hookrightarrow \bbE^{n}$ is minimal\footnote{The Euler-Lagrange
equation for Nambu action states that the map $\bX: \Sigma \to
\bbE^{n}$ is harmonic with respect to the induced metric.},
\emph{i.e.}, the mean curvature vector vanishes.  By substituting
\eqref{eq:hessian-X} into eq.~\eqref{eq:eom-normal} we obtain a second
order partial differential equation for the embedding map $\bX^{\mu}$
where the Lovelock tensor terms provide a quadratic form that mimics a
metric.

\begin{remark}
    A different derivation of \eqref{eq:eom-normal} in terms of more
    conventional tensor analysis is the following.  Use
    \eqref{eq:delta-g} in the form $\delta g_{ab} =
    (\partial_{a}X^{\mu})(\partial_{b}(\delta X^{\mu})) +
    (\partial_{a}(\delta X^{\mu}))(\partial_{b}X^{\mu})$, insert it
    into \eqref{eq:var-love-1}, integrate by parts, use the
    conservation of the Lovelock tensors, and substitute  kinematic
    result \eqref{eq:hessian-X}.
\end{remark}

\begin{remark}
    We also provide an alternative derivation of
    eq.~\eqref{eq:hessian-X} \emph{\`{a} la Cartan}.  We have a map
    $\bX: \Stil \to \bbE^{n}$ with $\bX(\sigma) = \bE_{\mu}\,
    X^{\mu}(\sigma)$. Therefore there exists functions
    $X^{\mu}{}_{a}$ on $\Stil$ such that $dX^{\mu} =
    X^{\mu}{}_{a}\,\theta^{a}$.  Next we see that $0 = d^{2}X^{\mu} =
    (dX^{\mu}{}_{a} + \omega_{ab}X^{\mu}{}_{b})\wedge \theta^{a} =
    DX^{\mu}{}_{a} \wedge \theta^{a}$.  Cartan's Lemma tells us that
    there exist functions $X^{\mu}{}_{ab} = X^{\mu}{}_{ba}$ on
    $\Stil$ such that $DX^{\mu}{}_{a} = X^{\mu}{}_{ab}\theta^{b}$.
    The $X^{\mu}{}_{ab}$ are the second covariant derivatives of
    $X^{\mu}$.  Next we use the kinematic relation
    $\ehat_{a}\,\theta^{a} = \bE_{\mu}\, dX^{\mu} = \bE_{\mu}\,
    X^{\mu}{}_{a} \theta^{a}$ to conclude that $\ehat_{a} =
    \bE_{\mu}\, X^{\mu}{}_{a}$.  Taking the exterior derivative of this
    last result and using \eqref{eq:def-de} we obtain $\ehat_{b}
    \omega_{ba} - \nhat_{j} K^{j}{}_{ab} = \bE_{\mu} dX^{\mu}{}_{a}$.
    Therefore we conclude that $\bE_{\mu}\left( dX^{\mu}{}_{a} +
    \omega_{ab} X^{\mu}{}_{b}\right) = - \nhat_{j} K^{j}{}_{ab}$ or
    equivalently $\bE_{\mu}\, X^{\mu}{}_{ab} = - \nhat_{j}\,
    K^{j}{}_{ab}$.
\end{remark}

We will not discuss the path integral measure in this article.  There
you have to study how the change of variables from the embedding
measure $[\mathcal{D}X]$ contribution in the path integral transforms
into the appropriate expression in terms of the induced metric
$\bigl([\mathcal{D}g]\, [\mathcal{D}(\text{other fields})]\;
\mathcal{J}\bigr)$, where $\mathcal{J}$ is the appropriate Jacobian.

\section{Topological defects}
\label{sec:defects}

We use topological defects as a model for the QFT that localizes the
energy density near a submanifold $\Sigma^{q}$.
D.~F\"{o}rster~\cite{Forster:1974ga} observed that, to leading order,
the effective action that describes the dynamics of a Nielsen-Olesen
vortex was the Nambu-Goto action for a string.  Years later, motivated
by cosmic strings, Maeda and Turok~\cite{Maeda:1987pd}, and
Gregory~\cite{Gregory:1988qv} computed the finite width corrections
for F\"{o}rster's results.  Some misconceptions concerning the
rigidity of cosmic strings in these works were clarified by Gregory,
Haws and Garfinkle~\cite{Gregory:1989gg}.  Later
Gregory~\cite{Gregory:1990pm} generalized these observations to
$p$-dimensional defects in gauge theories.  A synopsis of these works
is that if you have a $p$-dimensional defect ($q=p+1$) then the
effective action that governs the dynamics of the defect is of the
form
\begin{equation*}
    \int_{\Sigma} d^{q}\sigma\; \sqrt{-\det g_{\Sigma}\,} \left(a + b\;
    R_{\Sigma} + \dotsb\right),
\end{equation*}
where $a$ and $b$ are constants that depend on the explicit details of
the model.  The equations of motion for the defect will
be~\eqref{eq:eom-normal}.  Additionally, Gregory~\cite{Gregory:1990pm}
observed that to get a consistent expansion in the width of the
defect, she had to impose that the submanifold $\Sigma^{q}$ was
minimal.  In this Section we will reproduce and generalize the results
of the aforementioned authors results by applying the the Weyl volume
formulas.  One of the outcomes of this section, even though we work in
a specific model for expositional simplicity, is that the results are
very general and do not details of the QFT.

There are two distinct issues to consider in discussing the effective 
action for a defect:
\begin{enumerate}
    \item What is the defect worldbrane $\Sigma$?  How do you
    construct an approximate solution with defect worldbrane $\Sigma$?
    
    \item How good is this approximate solution you constructed?
\end{enumerate}
We discuss these below.

\subsection{Constructing an approximate solution}
\label{sec:approximate}

Assume we start with a Minkowski space field theory in $\bbM^{n}$
characterized by Lagrangian $\lag$ with fields (scalar, vector, etc.)
that we will simply denote by $\Phi$.  The action is invariant with
respect to the action of the Poincar\'{e} group $\Poin(n)$ on the
fields.  We will not be mathematically precise and we take the
following working definition.  A $p$-dimensional \emph{static defect}
is a topologically stable solution to the equations of motion that is
invariant with respect to the action of the subgroup $\Poin(q) \times
\SOrth(k) \subset \Poin(n)$ where $q=p+1$ and $1 \le q+k \le n$.
Notice that the invariance group of the solution implies that the
defect is static by definition.  When $k$ is maximal, $k = n-q =l$,
then the defect is said to be spherically symmetric.  We also assume
that the energy density of the fields $\Phi$ is localized relative to
the directions transverse to the defect.  The question addressed by
F\"{o}rster and subsequent researchers is ``What is the effective
action that governs the dynamics of a defect?''

Assume we take a static defect, denoted by $\phi_{0}$, and distort it
a little bit and let it evolve.  This dynamic solution to the
equations of motion will be denoted by $\Phi$.  The first remark is
that in symmetry breaking Higgs type models, the evolving defect will
also have a core.  A clarifying example is the abelian Higgs model.
Among the fields $\Phi$ there is a complex valued field $\varphi$ that
transforms as $\varphi \to e^{i\alpha}\;\varphi$ under the action of
the $\U(1)$ gauge group.  The core of the defect is located at the
codimension $2$ submanifold $\Sigma = \{\bx \in \bbM^{n} \;\mid\;
\varphi(\bx)=0\}$ whose existence is guaranteed by topological
considerations.  Note that condition $\varphi(\bx)=0$ is a gauge
invariant condition.

F\"{o}rster proposed a method for understanding the motion of the core
by deforming the static defect $\phi_{0}$ with a diffeomorphism $f$ of
Minkowski space.  The deformed defect $\Phi$ is specified by
$f^{*}\Phi = \Phi\circ f = \phi_{0}$.  The core $\Sigma_{0}= \bbM^{1+p}=
\bbM^{q} \subset \bbM^{n}$ of the static defect is mapped into the
core $\Sigma = f(\Sigma_{0})$ of $\Phi$.  In general, $\Phi$ will not be a
solution to the equations of motion but if the diffeomorphism is close
to the identity then $\Phi$ will be an approximate solution.
F\"{o}rster's proposal is to find the $\Sigma$ that gives the best
approximate solution and study the evolution of $\Sigma$.
In a more general model, we expect a similar formulation where the core is
the locus of points determined by some gauge invariant condition
imposed on the fields.  

Quantum field theories that interest us are diffeomorphism covariant.
Let $\bg$ is the metric, and let $\Phi$ collectively denote all the
fields. If $f$ is a diffeomorphism of spacetime then the action
satisfies the covariance requirement
\begin{equation}
    I(\Phi, \bg) = I(f^{*}\Phi, f^{*}\bg)\,
    \label{eq:def-covariance}
\end{equation}
where $f^{*}\Phi$ and $f^{*}\bg$ denotes the action of the
diffeomorphism on the fields and the metric respectively.

Let $\Sigma_{0}^{q} \subset \bbM^{n}$ be the standard time-like
$q$-plane with Minkowski coordinates $\sigma^{a}$, and let $\nu^{i}$
be cartesian coordinates normal to $\Sigma_{0}^{q}$; note that
$(\sigma,\nu)$ are standard coordinates on $\bbM^{n}$.  The
submanifold $\Sigma^{q}$ is the image of an embedding map $\bX:
\Sigma_{0}^{q} \to \Sigma^{q} \hookrightarrow \bbM^{n}$.  To construct
the coordinate system adapted to the tube that was used in our
computations, we require an extension of the map $\bX$ by choosing a
framing of the normal bundle of $\Sigma$: $\sigma\in \Sigma_{0}^{q}
\mapsto \bigl(\bX(\sigma) \in \Sigma^{q}, \nhat(\sigma) \in
T_{\sigma}\Sigma^{\perp} \bigr)$.  Such a map leads to a local
diffeomorphism $F_{\bX,\nhat}$ between tubular neighborhoods of
$\Sigma_{0}^{q}$ and $\Sigma^{q}$ given by eq.~\eqref{eq:def-bx}:
\begin{equation}
    F_{\bX,\nhat}: (\sigma,\nu) \mapsto \bX(\sigma) + \nu^{i}\nhat_{i}\,.
    \label{eq:def-F}
\end{equation}
Our static defect $\phi_{0}$, a solution to the equations of motion in
Minkowski space, has core $\Sigma_{0}^{q}$, and $\phi_{0}$ only depends
on the coordinates transverse to $\Sigma_{0}^{q}$.  Given the
diffeomorphism $F_{\bX,\nhat}$, we can construct a field configuration
$\Phi$ that is a deformation of the defect and specified by
$F_{\bX,\nhat}^{*}\Phi = \phi_{0}$.  Since $F_{\bX,\nhat}^{*}\Phi =
\Phi \circ F_{\bX,\nhat}$ we have that
\begin{equation}
    \Phi(\bx) = \Phi\bigl( \bX(\sigma) + \nu^{i}\nhat_{i} \bigr) =
    \phi_{0}(\nu^{1},\dotsc, \nu^{l}).
    \label{eq:def-FPhi}
\end{equation}
Let $\mink$ be the Minkowski metric on $\bbM^{n}$ then equation
\eqref{eq:def-F} and covariance of the action tells us that
\begin{equation}
    I(\Phi,\mink) = I(F_{\bX,\nhat}^{*}\Phi,F_{\bX,\nhat}^{*}\mink) = I(\phi_{0},F_{\bX,\nhat}^{*}\mink).
    \label{eq:action-trans}
\end{equation}
The metric $F_{\bX,\nhat}^{*}\mink$ is eq.~\eqref{eq:metric} but with Lorentzian 
signature.

Next we compute the action for the deformed 
defect by using the right hand side of \eqref{eq:action-trans}.  For
expositional simplicity, we ignore the vector fields and only
look at the scalar fields.  We observe that
\begin{align*}
    d\phi_{0} & = \frac{\partial \phi_{0}}{\partial \nu^{i}}\; 
    d\nu^{i}
    = (\partial_{i}\phi_{0})\, D\nu^{i} -
    (\partial_{i}\phi_{0})\,\omega^{ij} \nu^{j}\\
    &= (\partial_{i}\phi_{0})\, D\nu^{i} -
    (\partial_{i}\phi_{0})\,\Gamma_{a}{}^{ij} \nu^{j} \theta^{a}\\
    &= (\partial_{i}\phi_{0})\, D\nu^{i} -
    (\partial_{i}\phi_{0})\,\Gamma_{a}{}^{ij} \nu^{j}
    (I+\bnu\bdot\bK)_{ab}^{-1}\, \thhat^{b}\,.
\end{align*}
where $\omega^{ij} = \Gamma_{a}{}^{ij}\,\theta^{a}$. Since 
$(\thhat^{a},D\nu^{i})$ is an orthonormal coframe we have that the 
action for the scalar field becomes
\begin{subequations}\label{eq:action-0}
\begin{align}
    I(\phi_{0},F_{\bX,\nhat}^{*}\mink) &= -\int_{\Sigma} \dual_{\Sigma} \int
    d^{l}\nu\; \det(I + \bnu\bdot\bK) \nonumber\\
    &\quad\times
    \left[ \half h^{ab}(\sigma,\nu) \Gamma_{a}{}^{ij}(\sigma) 
    (\partial_{i}\phi_{0})\nu^{j}\; \Gamma_{b}^{kl}(\sigma) 
    (\partial_{k}\phi_{0})\nu^{l}
    +\half \delta^{ij} 
    (\partial_{i}\phi_{0})(\partial_{i}\phi_{0})\right]
    \nonumber \\
    &\quad - \int_{\Sigma} \dual_{\Sigma} \int d^{l}\nu\; 
    \det(I + \bnu\bdot\bK)\; V(\phi_{0})\,,
    \nonumber \\
    &= -\int_{\Sigma} \dual_{\Sigma} \int d^{l}\nu\; 
    \det(I + \bnu\bdot\bK)\;\left[ \half \delta^{ij}
    (\partial_{i}\phi_{0})(\partial_{i}\phi_{0}) + V(\phi_{0})\right] 
    \label{eq:action-a}\\
    -\int_{\Sigma} \dual_{\Sigma} \int d^{l}\nu\;
     & \det(I + \bnu\bdot\bK)\; \left[
    \half h^{ab}(\sigma,\nu) \Gamma_{a}{}^{ij}(\sigma) 
    (\partial_{i}\phi_{0})\nu^{j}\; \Gamma_{b}^{kl}(\sigma) 
    (\partial_{k}\phi_{0})\nu^{l} \right]\,.
    \label{eq:action-b}
\end{align}
\end{subequations}
In the above $h^{ab}$ is the inverse matrix of the metric in
eq.~\eqref{eq:metric-coords}.  Equations~\eqref{eq:action-0} are
general and do not depend on the assumptions of spherical symmetry.
Notice that summand \eqref{eq:action-a} is invariant with respect to
$\SO(l)$ gauge transformations of the normal bundle and the second
summand \eqref{eq:action-b} is in general not gauge invariant because
of the presence of the normal connection $\Gamma_{a}{}^{ij}$.  This
means that \eqref{eq:action-b} depends on the choice of coframing for
the normal bundle in general and is a ``torsional energy''
contribution\footnote{``Torsional'' is used in the sense of the
response to a torque.}.  We also note that the presence of the inverse
metric $h^{ab}$ in \eqref{eq:action-b} means that this term if of type
\eqref{eq:energy-2} that is not amenable to the Weyl simplification.

In the spherically symmetric case the value of the action
\eqref{eq:action-0} should be independent of the coframing and we 
would like to verify it.  In this
case we have that $\phi_{0} = \phi_{0}(\lVert \bnu \rVert)$ and
\begin{equation*}
    \frac{\partial \phi_{0}}{\partial \nu^{i}} = 
    \frac{\nu^{i}}{\lVert \bnu \rVert}\, \phi_{0}'(\lVert \bnu 
    \rVert)\,.
\end{equation*}
In eq.~\eqref{eq:action-b} we have a term $ \Gamma_{a}{}^{ij}
(\partial_{i}\phi_{0})\nu^{j} = \Gamma_{a}{}^{ij} 
(\partial_{i}\phi_{0})
\nu^{i}\nu^{j} \phi'_{0}(\lVert \bnu\rVert)/\lVert \bnu\rVert =0$ 
because $\Gamma_{a}{}^{ij}$ is antisymmetric under $i \leftrightarrow j$.
We automatically have that summand \eqref{eq:action-b} is zero and 
the action reduces to
\begin{equation}
    I(\phi_{0},F_{\bX,\nhat}^{*}\mink) = -\int_{\Sigma} \dual_{\Sigma} \int
    d^{l}\nu \; \det(I + \bnu\bdot\bK) \left[ \half
    \phi_{0}'(\lVert\bnu\rVert)^{2} + V(\phi_{0})\right].
    \label{eq:action-spherical}
\end{equation}
Eq.~\eqref{eq:action-spherical} is exactly of the form
\eqref{eq:total-energy} needed to apply the Weyl volume element
methods.  In this model, the moments will be constants and they determine
the coupling strength of each Lovelock term. The idea is to find a 
diffeomorphism $F_{\bX,\nhat}$ that minimizes the action.

We mention that if you have a spherically symmetric defect described 
by fields $\phi_{0}$, $A_{0}$, \emph{etc.}, then the
general result is that
\begin{equation}
    I(\phi_{0},F_{\bX,\nhat}^{*}\mink) = \int_{\Sigma} \dual_{\Sigma}
    \int d^{l}\nu \; \det(I + \bnu\bdot\bK) \;
    \mathcal{L}_{\perp}(\phi_{0},A_{0})\,,
    \label{eq:action-spherical-1}
\end{equation}
where $\mathcal{L}_{\perp}$ is the Lagrangian density for the normal 
bundle. Because $\mathcal{L}_{\perp}$ is spherically symmetric we can 
use the Weyl method and get a Lovelock type action.

If we don't have spherical symmetry then the form of the action is
more subtle.  You expect that that the answer should depend on the
choice of framing for the normal bundle.  Summand \eqref{eq:action-a}
is treatable by the Weyl method, the answer does not depend on the
framing, and it will lead to an energy density $u=u(\bnu)$ where
eq.~\eqref{eq:energy-total} is applicable.  The analysis of the second
summand \eqref{eq:action-b} is more complicated.  First we rewrite
\eqref{eq:action-b} in the form\footnote{This is a term of type 
\eqref{eq:energy-2} mentioned previously.}
\begin{equation}
   -\half\int_{\Sigma} \dual_{\Sigma}\; \Gamma_{a}{}^{ij}(\sigma)
   \Gamma_{b}^{kl}(\sigma) \int d^{l}\nu\; \det\bigl(I + 
   \bnu\bdot\bK(\sigma)\bigr)\;
    h^{ab} \bigl(\bnu\bdot\bK(\sigma)\bigr) \;\nu^{j}\nu^{l}\;
   (\partial_{i}\phi_{0}) (\partial_{k}\phi_{0}) \,.
    \label{eq:summand-nonsym}
\end{equation}
where we make explicitly clear that $h^{ab}$ is a function of
$\bnu\bdot\bK(\sigma)$.  The normal bundle integral
\begin{equation*}
    \int d^{l}\nu\; \det\bigl(I + 
       \bnu\bdot\bK(\sigma)\bigr)\;
	h^{ab} \bigl(\bnu\bdot\bK(\sigma)\bigr) \;\nu^{j}\nu^{l}\;
       (\partial_{i}\phi_{0}) (\partial_{k}\phi_{0})
\end{equation*}
transforms tensorially under changes of the normal bundle framing
$\nhat$.  Note that we expect eq.~\eqref{eq:summand-nonsym} to depend
on the choice of framing because of the presence of the normal bundle
connection $\Gamma_{a}^{ij}$.  Eq.~\eqref{eq:summand-nonsym} is not of
the type directly amenable to the results of this paper, see
\eqref{eq:energy-2}.  You have do a power series expansion of $h^{ab}$
in powers of $\bnu\bdot\bK$, see eq.~\eqref{eq:metric-coords}, and
also a power series expansion in $\bnu\bdot\bK$ of the determinant.
Subsequently you can perform the $\nu$ integrals to obtain moments
that will combine with the $K$ and $\Gamma$ factors.  The total action
\eqref{eq:action-0} has to be minimized with respect to both the
choice of embedding map $\bX$, the choice of normal bundle framing
$\nhat$, and variation of the field configuration $\phi_{0}$.

\subsection{How good is the approximate solution?}
\label{sec:how-good}

There are two types of variations we can perform on action
\eqref{eq:action-spherical}: we can vary the field or we can vary the
embedding of the submanifold $\Sigma$, see
Section~\ref{sec:embedding}.  The variation of the embedding leads to
the gravity-like field equations of motion \eqref{eq:eom-normal}.
There may have additional equations of motion that arise from varying
the field configuration and we turn to this next.

We vary action \eqref{eq:action-spherical} by varying the field
$\phi_{0}(\lVert\bnu\rVert) \to \phi_{0}(\lVert\bnu\rVert) +
(\delta\phi)(\sigma,\bnu)$ where the variation is by an arbitrary
function which is not necessarily spherically symmetric.  We require
the variation $\delta\phi$ to vanish on $\Sigma$ because the core of
the defect $\Sigma$ is kept fixed\footnote{We already discussed what
happens when the core is deformed.}.  Note that the equation of motion
for the defect is $-\partial^{i}\partial_{i}\phi_{0} + V'(\phi_{0})=0$
and $\partial_{i}\phi_{0}(\lVert\bnu\rVert) =
(\nu^{i}/\lVert\bnu\rVert) \phi_{0}'(\lVert\bnu\rVert)$.  The
variation of the action is given by
\begin{align}
    \delta I &= -\int_{\Sigma}
    \dual_{\Sigma} \int d^{l}\nu \; \det(I + \bnu\bdot\bK) \left[
    -\partial^{i}\partial_{i}\phi_{0} +
    V'(\phi_{0})\right] \bigl(\delta\phi(\sigma,\bnu)\bigr) \nonumber \\
    &\quad + \int_{\Sigma} \dual_{\Sigma} \int d^{l}\nu \; \det(I +
    \bnu\bdot\bK) \Tr\!\left( (I + \bnu\bdot\bK)^{-1} K^{i}\right)
    (\partial_{i}\phi_{0}) \bigl(\delta\phi(\sigma,\bnu)\bigr)
    \nonumber \\
    &=\int_{\Sigma} \dual_{\Sigma} \int d^{l}\nu \; \det(I +
    \bnu\bdot\bK) \Tr\!\left( (I + \bnu\bdot\bK)^{-1} K^{i}\right)
    (\partial_{i}\phi_{0}) \bigl(\delta\phi(\sigma,\bnu)\bigr)
    \nonumber \\
    &=\int_{\Sigma} \dual_{\Sigma} \int d^{l}\nu \; \det(I +
    \bnu\bdot\bK) \Tr\!\left( (I + \bnu\bdot\bK)^{-1}
    \bnu\bdot\bK\right)
    \frac{\phi_{0}'(\lVert\bnu\rVert)}{\lVert\bnu\rVert}
    \bigl(\delta\phi(\sigma,\bnu)\bigr) 
    \nonumber \\
    &= \left.  \frac{\partial}{\partial\lambda}
    \right\rvert_{\lambda=1} \int_{\Sigma} \dual_{\Sigma} \int
    d^{l}\nu \; \det(I + \lambda\bnu\bdot\bK)
    \frac{\phi_{0}'(\lVert\bnu\rVert)}{\lVert\bnu\rVert}
    \bigl(\delta\phi(\sigma,\bnu)\bigr)
    \label{eq:var-action}
\end{align}
Equation~\eqref{eq:var-action} is precisely the type that we can 
apply eq.~\eqref{eq:energy-total}. First we write down the multipole 
expansion for $\delta\phi(\sigma,\bnu)$:
\begin{equation}
    \delta\phi(\sigma,\bnu) = \sum_{j=0}^{\infty}\sum_{i_{1},\dotsc, i_{j}} 
    (\delta\phi)^{(j)}_{i_{1}\dotsm i_{j}}(\sigma, \lVert \bnu \rVert) \;
    \Yh^{j}_{i_{1} \dotsm u_{j}}(\bnuhat)\,.
    \label{eq:df-multi-1}
\end{equation}
Since $\phi_{0}$ is spherically symmetric we have
\begin{equation}
    \frac{\phi_{0}'(\lVert\bnu\rVert)}{\lVert\bnu\rVert}\,
    \delta\phi(\sigma,\bnu) = \sum_{j=0}^{\infty}\sum_{i_{1},\dotsc,
    i_{j}} \left(\frac{\phi_{0}'(\lVert\bnu\rVert)}{\lVert\bnu\rVert}
    \;(\delta\phi)^{(j)}_{i_{1}\dotsm i_{j}}(\sigma, \lVert \bnu
    \rVert) \right) \Yh^{j}_{i_{1} \dotsm u_{j}}(\bnuhat)\,.
    \label{eq:df-multi-2}
\end{equation}
We insert \eqref{eq:df-multi-2} into the 
definition~\eqref{eq:radial-mom-j} and obtain
\begin{equation}
    \begin{split}
    (\delta\mu)^{(j)}_{k_{1}\dotsm k_{j}}{}_{,2s+j}(\sigma) &=
    \int_{(T_{\sigma}\Sigma)^{\perp}} \lVert \bnu \rVert^{2s+j} \;
    \left(\frac{\phi_{0}'(\lVert\bnu\rVert)}{\lVert\bnu\rVert}
    \;(\delta\phi)^{(j)}_{k_{1}\dotsm k_{j}}(\sigma, \lVert \bnu
    \rVert) \right)\; d^{l}\nu \,, \\
    &= V_{l-1}(S^{l-1})\int_{0}^{\infty}d\nu\; \nu^{2s+j+l-1}\;
    \left(\frac{\phi_{0}'(\lVert\bnu\rVert)}{\lVert\bnu\rVert}
    \;(\delta\phi)^{(j)}_{k_{1}\dotsm k_{j}}(\sigma, \lVert \bnu
    \rVert) \right)\,.
    \end{split}
    \label{eq:radial-mom-df}
\end{equation}
Inserting this expression into \eqref{eq:energy-total} leads to the 
variation of the action
\begin{multline}
	\delta I  = \sum_{j=0}^{q} \sum_{s=0}^{\lfloor(q-j)/2\rfloor} 
	\frac{C_{2j+2s}}{2^{s}\; s!}  \; (2s+j) 	\\
	\times  \int_{\Sigma} 
	(\delta\mu)^{(j)}_{k_{1}\dotsm k_{j}}{}_{,2s+j}(\sigma) \;
	\kappa_{b_{1}}{}^{k_{1}} \wedge \dotsb \wedge
	\kappa_{b_{j}}{}^{k_{j}} \wedge
	\Omega_{a_{1}a_{2}} \wedge \dotsb \wedge \Omega_{a_{2s-1}a_{2s}}
	\wedge \dual^{b_{1}\dotsm b_{j} a_{1} \dotsm a_{2s}}\,.
     \label{eq:variation-total}
\end{multline}
The factor of $(2s+j)$ arises from the differentiation with respect to
$\lambda$ in \eqref{eq:var-action} and counts the degree of
homogeneity of each summand in \eqref{eq:variation-total}, where you
have degree $1$ homogeneity in $\kappa$ and degree $2$ homogeneity in
$\Omega$ to be compatible with the Gauss equation.  Notice that
obtaining a non-zero summand requires $2s+j \ge 1$, and therefore there
is no trouble from the $1/\lVert\bnu\rVert$ term in
\eqref{eq:radial-mom-df} because $\codim \Sigma = l \ge 1$.

Computing the variation we have
\begin{multline}
    \frac{\delta I}{\ (\delta\phi)^{(j)}_{k_{1}\dotsm
    k_{j}}(\sigma, \lVert \bnu \rVert)} = \mathcal{N}_{j}\;
    V_{l-1}(S^{l-1})
    \frac{\phi_{0}'(\lVert\bnu\rVert)}{\lVert\bnu\rVert}
    \sum_{s=0}^{\lfloor(q-j)/2\rfloor} \frac{C_{2j+2s}}{2^{s}\;
    s!} \; (2s+j) \; \nu^{2s+j+l-1}\; \\
    \cdot  \left(\kappa_{b_{1}}{}^{k_{1}} \wedge \dotsb \wedge
    \kappa_{b_{j}}{}^{k_{j}}\right)^{\text{ST}} \!\wedge
    \Omega_{a_{1}a_{2}} \wedge \dotsb \wedge \Omega_{a_{2s-1}a_{2s}}
    \wedge \dual^{b_{1}\dotsm b_{j} a_{1} \dotsm a_{2s}}\;.
    \label{eq:full-var-action}
\end{multline}
In the above, the superscript $\text{ST}$ means orthogonal projection
onto the symmetric traceless part on the $k$-indices, and
$\mathcal{N}_{j}$ is a normalization constant that depends on how one
normalizes symmetric traceless tensors and how one normalizes
differentiation with respect to an object with symmetric traceless
indices\footnote{The details of the normalization are not important in
what follows.}. Therefore, the equations of motion for the field 
$\phi_{0}$ become
\begin{equation}
    \left(\kappa_{b_{1}}{}^{k_{1}} \wedge \dotsb \wedge
    \kappa_{b_{j}}{}^{k_{j}}\right)^{\text{ST}} \!\wedge
    \Omega_{a_{1}a_{2}} \wedge \dotsb \wedge \Omega_{a_{2s-1}a_{2s}}
    \wedge \dual^{b_{1}\dotsm b_{j} a_{1} \dotsm a_{2s}} =0\,,
    \label{eq:eom-field}
\end{equation}
because \eqref{eq:full-var-action} has to be true for each value of
$\lVert \bnu \rVert$ for $2s + j \ge 1$.  This imposes a finite number
of constraints on the intrinsic geometry and extrinsic geometry of the
surface $\Sigma$ because $j \le q$ and $0 \le s \le
\lfloor(q-j)/2\rfloor$.  In particular, the leading term is given by
the first constraint ($s=0$ and $j=1$) which is the unique case with
$2s+j=1$ and leads to
\begin{equation}
    \kappa_{b}{}^{k}\wedge \dual^{b} =0\,.
    \label{eq:first-constraint}
\end{equation}
This is equivalent to the condition that the mean curvature vector
vanishes, $\mink^{ab}\, K_{ab}{}^{k}=0$, \emph{i.e.}, the submanifold
$\Sigma$ is minimal.  The next order $2s+j=2$ consist of two cases
$s=1$, $j=0$; and $s=0$, $j=2$.  The first gives the constraint $R=0$,
and the second leads to quadratic constraints on the extrinsic
curvature tensor: $\left(
K_{aa}{}^{k_{1}}K_{bb}{}^{k_{2}}-K_{ab}{}^{k_{1}}K_{ab}{}^{k_{2}}
\right)^{\text{ST}} =0$.  In general, the diffeomorphism deformed
field configuration will not be an exact solution because you cannot
satisfy all the geometrical constraints \eqref{eq:eom-field}.  For an
approximate solution, you should at least satisfy the first constraint
\eqref{eq:first-constraint}, that the submanifold $\Sigma^{q}$ be a
minimal surface, see~\cite{Gregory:1990pm}.

Now we summarize the work of Section~\ref{sec:embedding} and
Section~\ref{sec:defects}.  You conclude that to next to leading order
you would have to satisfy two sets of equations.  The first comes from
varying the embedding:
\begin{equation}
    \left[- \mu^{(0)}_{0} \, \mink^{ab}\, + \mu^{(0)}_{2}\, C_{2}
    \left( R^{ab} - \half \mink^{ab}\,\, R\right) \right]\;
    K_{ab}{}^{i}  =0\,.\qquad (s=0,1;j=0)
    \label{eq:eom-emb-x}
\end{equation}
The second comes from varying the fields in the QFT via diffeomorphism
(the order parameter is $2s+j$) while keeping the embedding fixed:
\begin{subequations}\label{eq:eom-emb}
\begin{align}
    \mink^{ab}\,K_{ab}{}^{i} & =0\,,\qquad (s=0;j=1)
    \label{eq:eom-emb-01}\\
    R & = 0\,,\qquad (s=1;j=0)
    \label{eq:eom-emb-10}\\
    \left( 
    K_{aa}{}^{k_{1}}K_{bb}{}^{k_{2}}-K_{ab}{}^{k_{1}}K_{ab}{}^{k_{2}} 
    \right)^{\text{ST}}
    &= 0\,.\qquad (s=0;j=2)
    \label{eq:eom-emb-02}
\end{align}
\end{subequations}

For the moment we will only discuss \eqref{eq:eom-emb-x} and
\eqref{eq:eom-emb-01} because we are only looking for approximate
solutions.  We leave a discussion of the two higher order conditions
\eqref{eq:eom-emb-10} and \eqref{eq:eom-emb-02} to the future.  The
variation of the field via \eqref{eq:eom-emb-01} tells us that
$\Sigma^{q}$ is minimal.  This agrees with \eqref{eq:eom-emb-x} if we
eliminate the Einstein tensor term by setting $\mu^{(0)}_{2}=0$.  In
this case the dynamics of the embedded submanifold is given by the
harmonic map condition or dalembertian condition $D^{a}\partial_{a}
\bX^{\mu}=0$, see \eqref{eq:hessian-X}.  For an emergent gravity-like
theory we would like $\mu^{(0)}_{2} \neq 0$ to get graviton-like
dynamical degrees of freedom.  Imposing the minimality condition
reduces \eqref{eq:eom-emb-x} to $R^{ab}\, K_{ab}{}^{i}=0$, but for our
purposes it is better to use the equivalent equation involving the
Einstein tensor $\love{2}^{ab}\, K_{ab}{}^{i}=0$ because we can invoke
our previous analysis.  We have a gravity-like theory that satisfies
$\love{2}^{ab} K_{ab}{}^{i}=0$ and the minimal submanifold condition
$\mink^{ab}\,K_{ab}{}^{i}=0$.  To address whether you get a \emph{bona
fide} emergent theory of gravity you need an extension of the
Burstin-Cartan-Janet-Schl\"{a}fly isometric embedding theorem to the
case of a minimal embedding.  We are not aware of such a theorem.  If
there is no extended embedding theorem then the theory will be
gravity-like because in the weak field approximation we have
\eqref{eq:def-chi}, and it appears you are effectively losing the trace
degree of freedom in the metric perturbations $\hpertb_{ab}$.  We
leave this as an open discussion.

\section{Numerical estimates}
\label{sec:numerology}

Next we discuss the size of the couplings in the effective action
\eqref{eq:gen-Weyl-x}.  First we assume that the couplings are
independent of $\sigma$ such as in the case of a spherically symmetric
defect.  If we think of $E^{(0)}$ as an effective dimensionless energy
or effective action entering a Boltzmann factor then the dimensions of
$u^{(0)}$ are $[u^{(0)}] = L^{-n} = M^{n}$.  For a static defect, the
energy density is translationally invariant along the defect and we
expect from dimensional analysis that $u^{(0)} \sim T_{q}/\corr^{l}$
where the $q$-dimensional energy density $T_{q}$ is associated with
the core with dimension $[T_{q}]=L^{-q}= M^{q}$, and $\corr$ is a
correlation length associated with the transverse directions.  In fact
$T_{q}$ is the $p$-brane tension as we will see.  From
eq.~\eqref{eq:def-mu2r-x} we see that $\mu_{2r}^{(0)} \sim
T_{q}\corr^{2r}$.  Since $\mu_{0}^{(0)}$ is the $p$-brane tension, we
immediately have its identification with $T_{q}$.  Note that the ratio
of successive couplings is always $\mu_{2r+2}^{(0)}/\mu_{2r}^{(0)}
\sim \corr^{2}$.  There is a dependence on $l$ through the
coefficients $C_{2r}$ that enter into our integration formula we have
not taken into account in our very rough estimates.  Applying this
result we see that the $q$-dimensional reciprocal Newtonian
gravitational constant $G_{q}^{-1} = (\MPl_{q})^{q-2} \sim
\mu_{2}^{(0)} \sim \corr^{2} T_{q}$.  The conventionally defined
cosmological constant $\Lambda_{q}$ is given by $T_{q} =\Lambda_{q}
G_{q}^{-1}$.  If $q=4$ then the experimentally measured value of the
cosmological constant is $\Lambda \sim 10^{-52}~\text{m}^{-2}$.  If we
take our formulas seriously we could obtain $\corr \sim
10^{26}~\text{m} \sim 10^{10}~\text{ly}$, this is the radius of the
observed universe. 

In Kaluza-Klein type scenarios, AADV obtain a relationship between the
4D Planck mass $\MPl_{4}$, the $(4+l)$-dimensional Planck scale
$\MPl_{4+l}$, and the compactification scale $r_{c}$ of the form
$(\MPl_{4})^{2} = r_{c}^{l}(\MPl_{l+4})^{2+l}$.  This is a little
different that the relationship we find in these energy tube theories
where $(\MPl_{4})^{2} = \corr^{2+l}\, m_{4+l}^{4+l}$, and $m_{4+l}$ is
the energy scale of the $n=4+l$ dimensional theory.  We cannot use a
higher dimensional Planck mass here because there is no higher
dimensional gravity in the embedding space.  Notice that there is an
offset of $2$ in the exponents on the right hand side of the
respective formulas; for example the formula for the $l=3$ case of the
AADV scenarios is formally the same as our formula in the $l=1$
scenario.  Due to this offset of $2$ in transverse dimensionality, the
results in this scenario will be much smaller than the numbers found
in \cite{ArkaniHamed:1998nn,ArkaniHamed:1998rs}. Our version of their 
formula is
\begin{equation*}
    \corr \sim 10^{32/(l+2)-19}~\text{m}\; \left( 
    \frac{1~\text{TeV}}{m_{4+l}}\right)^{(l+4)/(l+2)}\,.
\end{equation*}

\appendix

\section{Multilinear algebra miscellanea}
\label{sec:misc}

\subsection{Euclidean Space}
\label{sec:Euclid}

In this section, all indices $i$, $j$, $k$ and $l$ take values
$1,2,\dotsc,n$ and are associated with orthogonal cartesian coordinates
in $\bbE^{n}$.  The skew symmetric Kronecker symbol is defined by
\begin{align}
    \delta^{j_{1}j_{2}\dotsm j_{m}}_{i_{1}i_{2} \dotsm i_{m}} &= 
    \det
    \begin{pmatrix}    
	\delta^{j_{1}}{}_{i_{1}}& \delta^{j_{1}}{}_{i_{2}} & 
	\dotsm & \delta^{j_{1}}{}_{i_{m}} \\
	\delta^{j_{2}}{}_{i_{1}}& \delta^{j_{2}}{}_{i_{2}} & 
	\dotsm & \delta^{j_{2}}{}_{i_{m}} \\
	\hdotsfor{4} 
	\nonumber \\
	\delta^{j_{m}}{}_{i_{1}}& \delta^{j_{m}}{}_{i_{2}} & 
	\dotsm & \delta^{j_{m}}{}_{i_{m}} 
    \end{pmatrix} \\
    &= \frac{1}{(n-m)!}\; \epsilon_{i_{1}i_{2}\dotsm i_{m}
    k_{m+1}k_{m+2}\dotsm k_{n}} \epsilon^{j_{1}j_{2}\dotsm j_{m}
    k_{m+1}k_{m+2}\dotsm k_{n}}\,.
    \label{eq:skew-kronecker}
\end{align}
Note the trace relation
\begin{equation}
    \delta^{i_{1}i_{2}\dotsm i_{m}}_{i_{1}i_{2} \dotsm i_{m}} = 
    \frac{n!}{(n-m)!}\;.
    \label{eq:delta-trace}
\end{equation}

Let $\hodge$ be the Hodge duality operator on $\bbE^{n}$.  Consider an
orthonormal coframe $(\theta^{1},\theta^{2},\dotsc, \theta^{n})$ and
define the Hodge dual forms by
\begin{equation}
    \dual^{i_{1}i_{2}\dotsm i_{m}} = \hodge\left(\theta^{i_{1}} \wedge 
    \dotsb \wedge \theta^{i_{m}}\right) = \frac{1}{(n-m)!}\; 
    \epsilon^{i_{1}i_{2}\dotsm i_{m}}{}_{j_{m+1} \dotsm j_{n}}\; 
    \theta^{j_{m+1}} \wedge \dotsb \wedge \theta^{j_{n}}\;.
    \label{eq:hodge-dual}
\end{equation}
The volume element is $\dual = \hodge 1$. For us the key identity will 
be
\begin{equation}
    \left(\theta^{i_{1}} \wedge \dotsb \wedge \theta^{i_{m}}\right) 
    \wedge \dual_{j_{1}\dotsm j_{m}} = \delta^{i_{1}\dotsm 
    i_{m}}_{j_{1}\dotsm j_{m}} \; \dual\;.
    \label{eq:basic-identity}
\end{equation}

If $\bR = \frac{1}{4}R_{ijkl}\, (\theta^{i}\wedge \theta^{j}) \otimes
(\theta^{k}\wedge \theta^{l})$ is the Riemann curvature tensor of a
manifold $M$, $\dim M = n$,  then the Ricci
tensor is defined by $R_{jl} = R^{i}{}_{jil}$ and the scalar curvature is $R =
R^{j}{}_{j}$.  With these conventions we have
\begin{align}
    \delta^{j_{1}j_{2}}_{i_{1}i_{2}}\; R^{i_{1}i_{2}}{}_{j_{1}j_{2}}& 
    = 2R\,,
    \label{eq:foo-1}\\
    \delta^{j_{1}j_{2}j_{3}j_{4}}_{i_{1}i_{2}i_{3}i_{4}}\; 
    R^{i_{1}i_{2}}{}_{j_{1}j_{2}} R^{i_{3}i_{4}}{}_{j_{3}j_{4}} & = 
    4 \left(R^{2} - 4 R^{ij}R_{ij} + R^{ijkl}R_{ijkl} \right).
    \label{eq:goo}
\end{align}

Let $S$ be a symmetric matrix then
\begin{align*}
    \det(I + t S) &= \frac{1}{n!}\; \epsilon_{i_{1}i_{2}\dotsm 
    i_{n}} \epsilon^{j_{1}j_{2}\dotsm j_{n}} (I + 
    tS)^{i_{1}}{}_{j_{1}} (I+tS)^{i_{2}}{}_{j_{2}} \dotsm 
    (I+tS)^{i_{n}}{}_{j_{n}} \,,\\
    &= \frac{t^{n}}{n!} \; \epsilon_{i_{1}i_{2}\dotsm
    i_{n-1}i_{n}} \epsilon^{j_{1}j_{2}\dotsm j_{n-1}j_{n}}
    S^{i_{1}}{}_{j_{1}} S^{i_{2}}{}_{j_{2}} \dotsm
    S^{i_{n-1}}{}_{j_{n-1}} S^{i_{n}}{}_{j_{n}} \\
    &\quad + \frac{t^{n-1}}{n!} \binom{n}{1} \epsilon_{i_{1}i_{2}\dotsm
    i_{n-1}k_{n}} \epsilon^{j_{1}j_{2}\dotsm j_{n-1}k_{n}}
    S^{i_{1}}{}_{j_{1}} S^{i_{2}}{}_{j_{2}} \dotsm
    S^{i_{n-1}}{}_{j_{n-1}}  \\
    &\quad + \frac{t^{n-2}}{n!} \binom{n}{2}
    \epsilon_{i_{1}i_{2}\dotsm i_{n-2}k_{n-1}k_{n}}
    \epsilon^{j_{1}j_{2}\dotsm j_{n-2}k_{n-1}k_{n}}
    S^{i_{1}}{}_{j_{1}} S^{i_{2}}{}_{j_{2}} \dotsm
    S^{i_{n-2}}{}_{j_{n-2}} \\
    &\quad + \frac{t^{n-3}}{n!} \binom{n}{3}
    \epsilon_{i_{1}i_{2}\dotsm i_{n-3} k_{n-2}k_{n-1}k_{n}}
    \epsilon^{j_{1}j_{2}\dotsm i_{n-3} k_{n-2}k_{n-1}k_{n}}
    S^{i_{1}}{}_{j_{1}} S^{i_{2}}{}_{j_{2}} \dotsm
    S^{i_{n-3}}{}_{j_{n-3}} \\
    &\quad + \dotsb
     + \frac{t}{n!} \;\binom{n}{n-1}\;
    \epsilon_{i_{1}k_{2}\dotsm k_{n-1}k_{n}}
    \epsilon^{j_{1}k_{2}\dotsm k_{n-1}k_{n}}
    S^{i_{1}}{}_{j_{1}}   
     + 1\;, \\
    &= 1 + t \delta^{j}{}_{i}\; S^{i}{}_{j} + \frac{t^{2}}{2!} \;
    \delta^{j_{1}j_{2}}_{i_{1}i_{2}}\; S^{i_{1}}{}_{j_{1}}
    S^{i_{2}}{}_{j_{2}} + \frac{t^{3}}{3!}\;
    \delta^{j_{1}j_{2}j_{3}}_{i_{1}i_{2}i_{3}}\;
    S^{i_{1}}{}_{j_{1}} S^{i_{2}}{}_{j_{2}}S^{i_{3}}{}_{j_{3}} 
    + \dotsb \\
    &\quad + \frac{t^{n-1}}{(n-1)!}\;
    \delta^{j_{1}j_{2}\dotsm j_{n-1}}_{i_{1}i_{2}\dotsm i_{n-1}}\;
    S^{i_{1}}{}_{j_{1}} S^{i_{2}}{}_{j_{2}}\dotsm 
    S^{i_{n-1}}{}_{j_{n-1}} \\
    &\quad + \frac{t^{n}}{n!}\;
    \delta^{j_{1}j_{2}\dotsm j_{n}}_{i_{1}i_{2}\dotsm i_{n}}\;
    S^{i_{1}}{}_{j_{1}} S^{i_{2}}{}_{j_{2}}\dotsm 
    S^{i_{n}}{}_{j_{n}}\;.
\end{align*}
Summarizing we have the very useful result
\begin{equation}
    \det(I + t S) = \sum_{m=0}^{n} \frac{t^{m}}{m!} \; \delta^{j_{1}\dotsm 
    j_{m}}_{i_{1}\dotsm i_{m}}\; S^{i_{1}}{}_{j_{1}} 
    S^{i_{2}}{}_{j_{2}} \dotsm S^{i_{m}}{}_{j_{m}}\;.
    \label{eq:det-expansion}
\end{equation}
A check on the above is to note that if $S=I$ then $\det(I+tS) =
(1+t)^{n}$ and this agrees with \eqref{eq:det-expansion} if we use
\eqref{eq:delta-trace}.

\subsection{Some Hodge duality differences in Minkowski space}
\label{sec:Mink}

Here we assume the signature of the metric is $(\tau, 
+1,+1,\dotsc,+1)$ where $\tau = \pm 1$ depending on whether we are 
in Euclidean space $\bbE^{n}$ or Minkowski space $\bbM^{n}$. We 
choose the index range to be $1,2,3,\dotsc,n$ where the index value 
$1$ corresponds to time in $\bbM^{n}$. We choose the convention that 
$\epsilon_{123\dotsm n}=+1$. If we raise the indices then 
$\epsilon^{123\dotsm n}=\tau$. Equation~\eqref{eq:skew-kronecker} 
becomes
\begin{equation}
    \tau\, \delta^{j_{1}j_{2}\dotsm j_{m}}_{i_{1}i_{2} \dotsm i_{m}} =
    \frac{1}{(n-m)!}\; \epsilon_{i_{1}i_{2}\dotsm i_{m}
    k_{m+1}k_{m+2}\dotsm k_{n}} \epsilon^{j_{1}j_{2}\dotsm j_{m}
    k_{m+1}k_{m+2}\dotsm k_{n}}\,.
    \label{eq:mink-kronecker}
\end{equation}
The Hodge dual is still defined by eq.~\eqref{eq:hodge-dual} and
eq.~\eqref{eq:basic-identity} still holds.  We note that $\hodge 1 =
\dual$ and $\theta^{i}\wedge \dual_{j} = \delta^{i}{}_{j}\,\dual$.

Next we point out that if $T_{i_{1}\dotsm i_{m}}$ is a rank $m$ tensor
then the induced inner product is usually taken to be $\lVert T
\rVert^{2} = T_{i_{1}\dotsm i_{m}}T^{i_{1}\dotsm i_{m}}$.  When
dealing with differential forms $\omega = \frac{1}{m!}\,
\omega_{i_{1}\dotsm i_{m}} \theta^{i_{1}}\wedge \dotsb \wedge
\theta^{i_{m}}$ it is convenient to define a slightly different
normalization that is convenient for Hodge duality purposes
$\hnorm{\omega} = \frac{1}{m!} \lVert \omega \rVert^{2}$.  With this
convention we have the basic Hodge duality relation $\omega \wedge
\hodge \omega = \hnorm{\omega}\;\dual$.  Since duality takes spacelike
subspaces to timelike subspaces we know that $\hnorm{\omega} = \tau
\hnorm{\hodge\omega}$.  This is enough to show that
$\hodge\hodge\omega = \tau (-1)^{m(n-m)}\,\omega$.  First we note that
$ \hnorm{\hodge\omega}\,\dual = (\hodge\omega)\wedge
\hodge(\hodge\omega) = (-1)^{m(n-m)}(\hodge\hodge\omega)\wedge
(\hodge\omega)$ At the same time we have $\hnorm{\hodge\omega}\,\dual
= \tau \hnorm{\omega}\,\dual= \tau\omega\wedge \hodge \omega$.
Comparing expressions we find that $\hodge\!\hodge \omega = \tau
(-1)^{m(n-m)}\,\omega$.  Finally, we remark that $\hodge \dual =
\tau\cdot 1$, and $\hodge\left( \dual^{i_{1}\dotsm i_{m}}\right) =
\tau (-1)^{m(n-m)}\; \theta^{i_{1}} \wedge \dotsb\wedge
\theta^{i_{m}}$.

\section{Averaging over a sphere}
\label{sec:avg-sphere}

Let $S^{d-1} \subset \bbE^{d}$ then the $(d-1)$-volume of $S^{d-1}$ 
is 
\begin{equation}
    V_{d-1}(S^{d-1}) = \frac{2 \pi^{d/2}}{\Gamma(d/2)}\,. 
    \label{eq:volume-sphere}
\end{equation}
The unit $d$-ball, \emph{a.k.a.}, the solid sphere or the disk, is
$B^{d} = \{ \bx\in \bbE^{n}\;\mid\; \lVert \bx \rVert \le 1\}$; note
that $\partial B^{d} = S^{d-1}$.  By integrating the volume element
over concentric shells it is easy to see that the $d$-volume of the
$d$-ball of radius $\rho$ is given by
\begin{equation}
    V_{d}(B^{d}_{\rho}) = \frac{V_{d-1}(S^{d-1})}{d}\; \rho^{d}
    = \frac{\pi^{d/2}}{\Gamma(d/2+1)}\, \rho^{d}\;.
    \label{eq:vol-ball}
\end{equation}

We define the average
\begin{equation}
    \VEV{x^{i_{1}}x^{i_{2}} \dotsm x^{i_{r}}} = 
    \frac{1}{V_{d-1}(S^{d-1})}\; \int_{S^{d-1}} x^{i_{1}}x^{i_{2}} \dotsm 
    x^{i_{r}} \; d\vol_{S^{d-1}}\;,
    \label{eq:def-VEV}
\end{equation}
where $d\vol_{S^{d-1}}$ is the induced volume element on the sphere by
the embedding $S^{d-1} \hookrightarrow \bbE^{d}$. If $r$ is an odd 
integer then the average vanishes by parity. For $r=2n$, consider 
the set of indices $\{i_{1},i_{2}, \dotsc, i_{2n}\}$ and let 
$\sym^{i_{1}i_{2}\dotsm i_{2n}}$ be a tensorial expression consisting 
of $(2n)!/2^{n} n! = (2n-1)!!$ monomials constructed out of Kronecker 
deltas by considering all possible ``Wick contractions'' of the 
indices. For example
\begin{equation}
    \begin{split}
        \sym^{ij} &= \delta^{ij}\;, \\
        \sym^{ijkl} &= \delta^{ij}\delta^{kl} + \delta^{ik}\delta^{jl} 
        + \delta^{il}\delta^{jk}\;,\\
        \vdots\quad &= \qquad\vdots \\
        \sym^{i_{1}i_{2}\dotsm i_{2n}} &= \underbrace{\delta^{i_{1}i_{2}} 
        \delta^{i_{3}i_{4}} \dotsm \delta^{i_{2n-1}i_{2n}} + 
        \dotsb}_{(2n-1)!! \text{ monomials}} \;.
    \end{split}
    \label{eq:wick}
\end{equation}
Note that $\sym$ is invariant under any permutation of the indices.

The basic averaging theorem is
\begin{equation}
    \VEV{x^{i_{1}}x^{i_{2}} \dotsm x^{i_{2n}}} = C_{2n}\, 
    \sym^{i_{1}i_{2}\dotsm i_{2n}}\;,
    \label{eq:sphere-avg}
\end{equation}
where $C_{0}=1$ , \emph{i.e.}, $\VEV{1}=1$; and
\begin{equation}
    C_{2n} = \prod_{k=0}^{n-1} \frac{1}{d+2k} \qquad\text{for $n \ge1$.}
    \label{eq:def-C}
\end{equation}
This theorem is a consequence of the theory of invariants for the
orthogonal group.  The normalization factor is easily obtained by 
setting $i_{i} = i_{2}$ in the formula \eqref{eq:sphere-avg}, summing
over $i_{1}$, applying the condition $x^{i_{1}}x^{i_{1}} =\lVert x
\rVert^{2} = 1$ and thus obtaining the recursion relation
$C_{2n+2}=C_{2n}/(d+2n)$.

\acknowledgments

I would like to thank Matthew Haddad, James Nearing, Rafael Nepomechie
and Paul Windey for discussions.  I would also like to thank John Lott
for suggesting some very helpful mathematical references.  K.~Alvarez
helped with the figures.  This work was supported in part by the
National Science Foundation under Grant PHY-1212337.

\relax

\def\cprime{$'$}
\providecommand{\href}[2]{#2}\begingroup\raggedright\endgroup

\end{document}

%% file: emer-grav-defs.tex
\newcommand{\VEV}[1]{\left\langle #1\right\rangle}

\newcommand*{\thhat}{\hat{\theta}}
\newcommand*{\lag}{\mathcal{L}}
\newcommand*{\br}{\bm{r}}
\newcommand*{\bx}{\bm{x}}

\newcommand*{\bnu}{\bm{\nu}}
\newcommand*{\bnuhat}{\bm{\hat{\nu}}}
\newcommand*{\nuhat}{\hat{\nu}}
\newcommand*{\evec}{\bm{e}}
\newcommand*{\ehat}{\bm{\hat{e}}}
\newcommand*{\nhat}{\bm{\hat{n}}}
\newcommand*{\bdot}{\bm{\cdot}}
\newcommand*{\bE}{\bm{\hat{E}}}
\newcommand*{\bX}{\bm{X}}
\newcommand*{\bK}{\bm{K}}
\newcommand*{\bgtil}{\bm{\tilde{g}}}
\newcommand*{\bbE}{\mathbb{E}}
\newcommand*{\bbM}{\mathbb{M}}

\newcommand*{\hodge}{\star}
\newcommand*{\sym}{\mathcal{W}}
\newcommand*{\tube}{\mathcal{T}}
\newcommand*{\curv}{\mathcal{K}}
\newcommand*{\cD}{\mathcal{D}}
\newcommand*{\Stil}{\widetilde{\Sigma}}
\newcommand*{\Yh}{\mathcal{Y}}
\newcommand*{\curvform}{\kappa}

\DeclareMathOperator{\Lie}{\mathcal{L}}
\DeclareMathOperator{\Sym}{Sym}

\DeclareMathOperator*{\Tr}{Tr}
\DeclareMathOperator{\Diff}{Diff}
\DeclareMathOperator{\Map}{Map}

\DeclareMathOperator{\codim}{codim}

\DeclareMathOperator{\SO}{SO}

\DeclareMathOperator{\pf}{pf}
\DeclareMathOperator{\rank}{rank}
\DeclareMathOperator{\vol}{vol}

\DeclareMathOperator{\SOrth}{SO}
\DeclareMathOperator{\U}{U}
\DeclareMathOperator{\Poin}{\mathcal{P}}

\newcommand*{\half}{\frac{1}{2}}

\newcommand*{\mink}{\bm{g}_{\bbM}}
\newcommand*{\eucl}{\bm{g}_{\bbE}}
\newcommand*{\bg}{\bm{g}}
\newcommand*{\corr}{\xi_{\perp}}
\newcommand*{\bR}{\bm{R}}
\newcommand{\hnorm}[1]{\left\lVert #1 \right\rVert^{2}_{\hodge}}

\newcommand{\love}[1]{E_{[#1]}}
\newcommand*{\dual}{\bm{\zeta}}
\newcommand*{\MPl}{M^{\text{Pl}}}
\DeclareMathOperator{\AdS}{AdS}
\newcommand*{\hpert}{\gamma}
\newcommand*{\hpertb}{\bar{\gamma}}

%% file: emergent-gravity.bbl
\begin{thebibliography}{10}

\bibitem{Dai:1989ua}
J.~Dai, R.~G. Leigh and J.~Polchinski, \emph{{New Connections Between String
  Theories}}, \href{http://dx.doi.org/10.1142/S0217732389002331}{\emph{Mod.
  Phys. Lett.} {\bf A4} (1989) 2073--2083}.

\bibitem{Antoniadis:1998ig}
I.~Antoniadis, N.~Arkani-Hamed, S.~Dimopoulos and G.~Dvali, \emph{{New
  dimensions at a millimeter to a Fermi and superstrings at a TeV}},
  \href{http://dx.doi.org/10.1016/S0370-2693(98)00860-0}{\emph{Phys.Lett.} {\bf
  B436} (1998) 257--263}, [\href{http://arxiv.org/abs/hep-ph/9804398}{{\tt
  hep-ph/9804398}}].

\bibitem{Forster:1974ga}
D.~Forster, \emph{{Dynamics of Relativistic Vortex Lines and their Relation to
  Dual Theory}},
  \href{http://dx.doi.org/10.1016/0550-3213(74)90008-X}{\emph{Nucl.Phys.} {\bf
  B81} (1974) 84}.

\bibitem{Maeda:1987pd}
K.-i. Maeda and N.~Turok, \emph{{Finite Width Corrections to the Nambu Action
  for the Nielsen-Olesen String}},
  \href{http://dx.doi.org/10.1016/0370-2693(88)90488-1}{\emph{Phys.Lett.} {\bf
  B202} (1988) 376}.

\bibitem{Gregory:1988qv}
R.~Gregory, \emph{Effective action for a cosmic string},
  \href{http://dx.doi.org/10.1016/0370-2693(88)91492-X}{\emph{Phys.Lett.} {\bf
  B206} (1988) 199}.

\bibitem{Weyl:tubes}
H.~Weyl, \emph{On the {V}olume of {T}ubes},
  \href{http://dx.doi.org/10.2307/2371513}{\emph{Amer. J. Math.} {\bf 61}
  (1939) 461--472}.

\bibitem{Lovelock:1971yv}
D.~Lovelock, \emph{{The Einstein tensor and its generalizations}},
  \href{http://dx.doi.org/10.1063/1.1665613}{\emph{J.Math.Phys.} {\bf 12}
  (1971) 498--501}.

\bibitem{Lovelock:1972vz}
D.~Lovelock, \emph{{The four-dimensionality of space and the Einstein tensor}},
  \href{http://dx.doi.org/10.1063/1.1666069}{\emph{J.Math.Phys.} {\bf 13}
  (1972) 874--876}.

\bibitem{Zwiebach:1985uq}
B.~Zwiebach, \emph{{Curvature Squared Terms and String Theories}},
  \href{http://dx.doi.org/10.1016/0370-2693(85)91616-8}{\emph{Phys.Lett.} {\bf
  B156} (1985) 315}.

\bibitem{Zumino:1985dp}
B.~Zumino, \emph{{Gravity Theories in More Than Four-Dimensions}},
  \href{http://dx.doi.org/10.1016/0370-1573(86)90076-1}{\emph{Phys.Rept.} {\bf
  137} (1986) 109}.

\bibitem{Maldacena:1997re}
J.~M. Maldacena, \emph{{The Large N limit of superconformal field theories and
  supergravity}}, \href{http://dx.doi.org/10.1023/A:1026654312961,
  10.1023/A:1026654312961}{\emph{Adv.Theor.Math.Phys.} {\bf 2} (1998)
  231--252}, [\href{http://arxiv.org/abs/hep-th/9711200}{{\tt
  hep-th/9711200}}].

\bibitem{Aharony:1999ti}
O.~Aharony, S.~S. Gubser, J.~M. Maldacena, H.~Ooguri and Y.~Oz, \emph{{Large N
  field theories, string theory and gravity}},
  \href{http://dx.doi.org/10.1016/S0370-1573(99)00083-6}{\emph{Phys.Rept.} {\bf
  323} (2000) 183--386}, [\href{http://arxiv.org/abs/hep-th/9905111}{{\tt
  hep-th/9905111}}].

\bibitem{Witten:1988xj}
E.~Witten, \emph{{Topological Sigma Models}},
  \href{http://dx.doi.org/10.1007/BF01466725}{\emph{Commun. Math. Phys.} {\bf
  118} (1988) 411}.

\bibitem{Witten:1988ze}
E.~Witten, \emph{{Topological Quantum Field Theory}},
  \href{http://dx.doi.org/10.1007/BF01223371}{\emph{Commun. Math. Phys.} {\bf
  117} (1988) 353}.

\bibitem{Bershadsky:1993cx}
M.~Bershadsky, S.~Cecotti, H.~Ooguri and C.~Vafa, \emph{{Kodaira-Spencer theory
  of gravity and exact results for quantum string amplitudes}},
  \href{http://dx.doi.org/10.1007/BF02099774}{\emph{Commun. Math. Phys.} {\bf
  165} (1994) 311--428}, [\href{http://arxiv.org/abs/hep-th/9309140}{{\tt
  hep-th/9309140}}].

\bibitem{ArkaniHamed:1998nn}
N.~Arkani-Hamed, S.~Dimopoulos and G.~Dvali, \emph{{Phenomenology, astrophysics
  and cosmology of theories with submillimeter dimensions and TeV scale quantum
  gravity}},
  \href{http://dx.doi.org/10.1103/PhysRevD.59.086004}{\emph{Phys.Rev.} {\bf
  D59} (1999) 086004}, [\href{http://arxiv.org/abs/hep-ph/9807344}{{\tt
  hep-ph/9807344}}].

\bibitem{ArkaniHamed:1998rs}
N.~Arkani-Hamed, S.~Dimopoulos and G.~Dvali, \emph{{The Hierarchy problem and
  new dimensions at a millimeter}},
  \href{http://dx.doi.org/10.1016/S0370-2693(98)00466-3}{\emph{Phys.Lett.} {\bf
  B429} (1998) 263--272}, [\href{http://arxiv.org/abs/hep-ph/9803315}{{\tt
  hep-ph/9803315}}].

\bibitem{Randall:1999vf}
L.~Randall and R.~Sundrum, \emph{{An Alternative to compactification}},
  \href{http://dx.doi.org/10.1103/PhysRevLett.83.4690}{\emph{Phys.Rev.Lett.}
  {\bf 83} (1999) 4690--4693}, [\href{http://arxiv.org/abs/hep-th/9906064}{{\tt
  hep-th/9906064}}].

\bibitem{Randall:1999ee}
L.~Randall and R.~Sundrum, \emph{{A Large mass hierarchy from a small extra
  dimension}},
  \href{http://dx.doi.org/10.1103/PhysRevLett.83.3370}{\emph{Phys.Rev.Lett.}
  {\bf 83} (1999) 3370--3373}, [\href{http://arxiv.org/abs/hep-ph/9905221}{{\tt
  hep-ph/9905221}}].

\bibitem{Alv:2016b}
O.~Alvarez and M.~J. Haddad, ``Emergent gravity in spaces of constant
  curvature.'', submitted for publication.

\bibitem{Dvali:2000hr}
G.~R. Dvali, G.~Gabadadze and M.~Porrati, \emph{{4-D gravity on a brane in 5-D
  Minkowski space}},
  \href{http://dx.doi.org/10.1016/S0370-2693(00)00669-9}{\emph{Phys. Lett.}
  {\bf B485} (2000) 208--214}, [\href{http://arxiv.org/abs/hep-th/0005016}{{\tt
  hep-th/0005016}}].

\bibitem{Gray:tubes}
A.~Gray, \emph{Tubes}, vol.~221 of \emph{Progress in Mathematics}.
\newblock Birkh\"auser Verlag, Basel, second~ed., 2004,
  \href{http://dx.doi.org/10.1007/978-3-0348-7966-8}{10.1007/978-3-0348-7966-8}.

\bibitem{Cartan:riemann}
{\'E}.~Cartan, \emph{Le\c cons sur la {G}\'eom\'etrie des {E}spaces de
  {R}iemann}.
\newblock Gauthier-Villars, Paris, 1946.

\bibitem{BCG3}
R.~L. Bryant, S.~S. Chern, R.~B. Gardner, H.~L. Goldschmidt and P.~A.
  Griffiths, \emph{Exterior differential systems}, vol.~18 of
  \emph{Mathematical Sciences Research Institute Publications}.
\newblock Springer-Verlag, New York, 1991,
  \href{http://dx.doi.org/10.1007/978-1-4613-9714-4}{10.1007/978-1-4613-9714-4}.

\bibitem{Johnson:D-branes}
C.~V. Johnson, \emph{D-branes}.
\newblock Cambridge University Press, New York, 2003.

\bibitem{FultonHarris}
W.~Fulton and J.~Harris, \emph{Representation theory}, vol.~129 of
  \emph{Graduate Texts in Mathematics}.
\newblock Springer-Verlag, New York, 1991,
  \href{http://dx.doi.org/10.1007/978-1-4612-0979-9}{10.1007/978-1-4612-0979-9}.

\bibitem{Greene:isometric}
R.~E. Greene, \emph{Isometric embeddings of {R}iemannian and
  pseudo-{R}iemannian manifolds.}
\newblock Memoirs of the American Mathematical Society, No. 97. American
  Mathematical Society, Providence, R.I., 1970.

\bibitem{Clarke:embedding}
C.~J.~S. Clarke, \emph{On the global isometric embedding of pseudo-{R}iemannian
  manifolds}, \href{http://dx.doi.org/10.1098/rspa.1970.0015}{\emph{Proc. Roy.
  Soc. London Ser. A} {\bf 314} (1970) 417--428}.

\bibitem{Berger:PNAS}
E.~Berger, R.~Bryant and P.~Griffiths, \emph{Some isometric embedding and
  rigidity results for {R}iemannian manifolds}, {\emph{Proc. Nat. Acad. Sci.
  U.S.A.} {\bf 78} (1981) 4657--4660}.

\bibitem{Bryant-Griffiths-Yang:isometric}
R.~L. Bryant, P.~A. Griffiths and D.~Yang, \emph{Characteristics and existence
  of isometric embeddings},
  \href{http://dx.doi.org/10.1215/S0012-7094-83-05040-8}{\emph{Duke Math. J.}
  {\bf 50} (1983) 893--994}.

\bibitem{Berger:big}
E.~Berger, R.~Bryant and P.~Griffiths, \emph{The {G}auss equations and rigidity
  of isometric embeddings},
  \href{http://dx.doi.org/10.1215/S0012-7094-83-05039-1}{\emph{Duke Math. J.}
  {\bf 50} (1983) 803--892}.

\bibitem{Eisenhart:RG}
L.~P. Eisenhart, \emph{Riemannian geometry}.
\newblock Princeton University Press, Princeton, NJ, 1966.

\bibitem{Gregory:1989gg}
R.~Gregory, D.~Haws and D.~Garfinkle, \emph{The dynamics of domain walls and
  strings}, \href{http://dx.doi.org/10.1103/PhysRevD.42.343}{\emph{Phys.Rev.}
  {\bf D42} (1990) 343--348}.

\bibitem{Gregory:1990pm}
R.~Gregory, \emph{{Effective actions for bosonic topological defects}},
  \href{http://dx.doi.org/10.1103/PhysRevD.43.520}{\emph{Phys.Rev.} {\bf D43}
  (1991) 520--525}.

\end{thebibliography}
